\documentclass{article}

\usepackage{arxiv}
\usepackage[round]{natbib}
\usepackage{subcaption,caption}

\usepackage[utf8]{inputenc} % allow utf-8 input
\usepackage[T1]{fontenc}    % use 8-bit T1 fonts
\usepackage{hyperref}       % hyperlinks
\usepackage{url}            % simple URL typesetting
\usepackage{booktabs}       % professional-quality tables
\usepackage{amsfonts}       % blackboard math symbols
\usepackage{nicefrac}       % compact symbols for 1/2, etc.
\usepackage{microtype}      % microtypography
\usepackage{lipsum}
\usepackage{graphicx}
\hypersetup{
    colorlinks=false,
    pdfborder={0 0 0},
    pdfborderstyle={/S/U/W 0},
}

\author{Travis Greene\\
	Institute of Service Science\\
	National Tsing Hua University, Taiwan\\
    \texttt{travis.greene@iss.nthu.edu.tw}\\
	\And
	Sofie Goethals\\
	Dept. of Engineering Management\\
	University of Antwerp, Belgium\\
    \texttt{sofie.goethals@uantwerpen.be}\\
	\And
	David Martens\\
	Dept. of Engineering Management\\
	University of Antwerp, Belgium\\
 \texttt{david.martens@uantwerpen.be}\\
	\And
	Galit Shmueli\\
	Institute of Service Science\\
	National Tsing Hua University, Taiwan\\
 \texttt{galit.shmueli@iss.nthu.edu.tw}
}

\begin{document}
\title{Algorithmic Explanations as Ad Opportunities:\\A Double-edged Sword}
\maketitle


\vspace*{5mm}
\begin{abstract}
Algorithms used by organizations increasingly wield power in society as they decide the allocation of key resources and basic goods. In order to promote fairer, juster, and more transparent uses of such decision-making power, explainable artificial intelligence (XAI) aims to provide %actionable 
insights into the logic of algorithmic decision-making. Despite much research on the topic, consumer-facing applications of XAI remain rare. A central reason may be that a viable platform-based monetization strategy for this new technology has yet to be found. We introduce and describe a novel monetization strategy for fusing algorithmic explanations with programmatic advertising via an \emph{explanation platform}. We claim the explanation platform represents a new, socially-impactful, and profitable form of human-algorithm interaction and estimate its potential for revenue generation in the high-risk domains of finance, hiring, and education. We then consider possible undesirable and unintended effects of monetizing XAI and simulate these scenarios using real-world credit lending data. Ultimately, we argue that monetizing XAI may be a double-edged sword: while monetization may incentivize industry adoption of XAI in a variety of consumer applications, it may also conflict with the original legal and ethical justifications for developing XAI. We conclude by discussing whether there may be ways to responsibly and democratically harness the potential of monetized XAI to provide greater consumer access to algorithmic explanations. 
\end{abstract}

\section{Introduction}
As humans, we naturally desire to understand and explain the world around us \citep{lombrozo2006structure}. This desire to understand extends to the social world as well. In our interactions with others, we seek causal knowledge of the mechanisms responsible for the actions of others \citep{gopnik2004theory}.  When we understand the reasons why others treat us as they do, we are more likely to trust and cooperate with them, two essential ingredients for successful collective action and a flourishing democracy \citep{hardin1999democracy}. The process and practice of explaining and justifying our decisions to others is a form of respect for persons \citep{rawls1971theory} and may have even fueled the evolution of the human capacity for reasoning itself \citep{mercier2011humans}.

But not all explanations are equally valid. Persons and organizations, especially those in positions of power, are expected to give appropriate reasons for their decisions, if those decisions are to be accepted as morally legitimate \citep{lazar2022legitimacy, locke1847essay}. This expectation remains relevant in the increasingly black-box digital world on which we depend for a variety of goods, services, and life opportunities \citep{pasquale2015black}. Access to meaningful explanations for algorithmic decisions is crucial if we are to trust organizations and new technologies with the power to make decisions affecting human well-being \citep{gabriel2022justiceAI}. Protecting the ethical and social value of explanations is consequently part of the task of developing \emph{human-centric AI} \citep{smuha2021race, shneiderman2020bridging}.

In light of these ethical and political ends, the European Union's (EU) General Data Protection Regulation (GDPR) gives persons rights to explanations for certain high-stakes algorithmic decisions affecting them \citep{gdpr}. The GDPR has the dual goals of protecting fundamental human rights and encouraging new forms of digital innovation \citep{albrecht2016gdpr}. More recently, the European Commission's proposed AI Act extends the democratic values of the GDPR by adding requirements of explanatory transparency for AI systems impacting access to public and private services and one's ability to ``fully participate in society'' \citep{NewEUAIreg2021}. And not to be left behind, in 2022 the United States' (US) White House Office of Science and Technology Policy released a Blueprint for an AI Bill of Rights, which includes provisions for algorithmic explanations, citing the need to protect democratic values and citizens' rights in the face of increasing algorithmic decision-making in society \citep{2022AibillWhitehouse}. The overarching aim of such legislation is to ``establish the governance infrastructure needed to hold bad actors accountable and allow actors with good intent to ensure and demonstrate that their [algorithms] are ethical, legal and safe'' \citep{mokander2022usAIA}.

Motivated in part by the GDPR, the field of explainable artificial intelligence (XAI) has witnessed an explosion of research attention. Along with preventing harm, being fair, and respecting human autonomy, XAI is likely to play a key role in the quest for responsible and trustworthy AI \citep{mittelstadt2019principles,martens2022dse}. XAI offers an exciting technological means of ensuring the democratic values of accountability and transparency guide the interactions of organizations, humans, and algorithms. XAI can illuminate the opaque algorithmic processes increasingly used for resume screening, platform work scheduling, credit scoring, distributing government benefits, and university scholarship and admissions decisions \citep{FPFAutoDecisionCourts2022}. With the tools of XAI, consumers can better understand complex algorithmic decisions and take actions that improve their chances of getting a positive decision in the future \citep{karimi2021surveyrecourse}. XAI thus promises considerable individual and social benefits. While certain public sector domains, such as healthcare and criminal justice, are also poised to benefit from XAI, we focus on three high-risk, consumer-facing industries that increasingly rely on automated decision-making: finance, education, and hiring.

%by offering them the causal knowledge necessary to

Despite mounting regulatory pressure and an ever-growing stream of XAI-related research, practical consumer-facing applications of XAI remain rare. A key reason may be that organizations have not yet identified a suitable monetization strategy. Although advertising has traditionally been a key monetization strategy for search engines, social media and other digital platforms, the synergies of XAI and platform-based digital advertising have not been investigated in the management or AI literature, despite the fact the US digital advertising market is growing and is expected to generate over \$200 billion in revenues by 2025 \citep{pwcInternetAds2022}. Much of this growth is due to advances in real-time-bidding (RTB) systems \citep{wang2017display} that allow individual-level ad impressions to be purchased by advertisers \citep{choi2020online}. Historically the infusion of advertising dollars has led to an explosion of innovative Web-based products and services \citep{hwang2020subprime, jansen2008sponsored}. Even major ride-sharing apps and platforms such as Uber and Lyft have taken steps to monetize their immense stores of user geolocation data to serve ads relevant to users' travel destinations and food delivery preferences \citep{uberlocationadsGizmodo2022}. We therefore think it reasonable to extrapolate from these trends and make the following claim: once consumer-facing XAI is appropriately monetized, it is likely to spread quickly through a number of different industries, sparking further technological and business innovation.

\begin{figure}[h]
         \centering
         \includegraphics[width=0.8\textwidth]{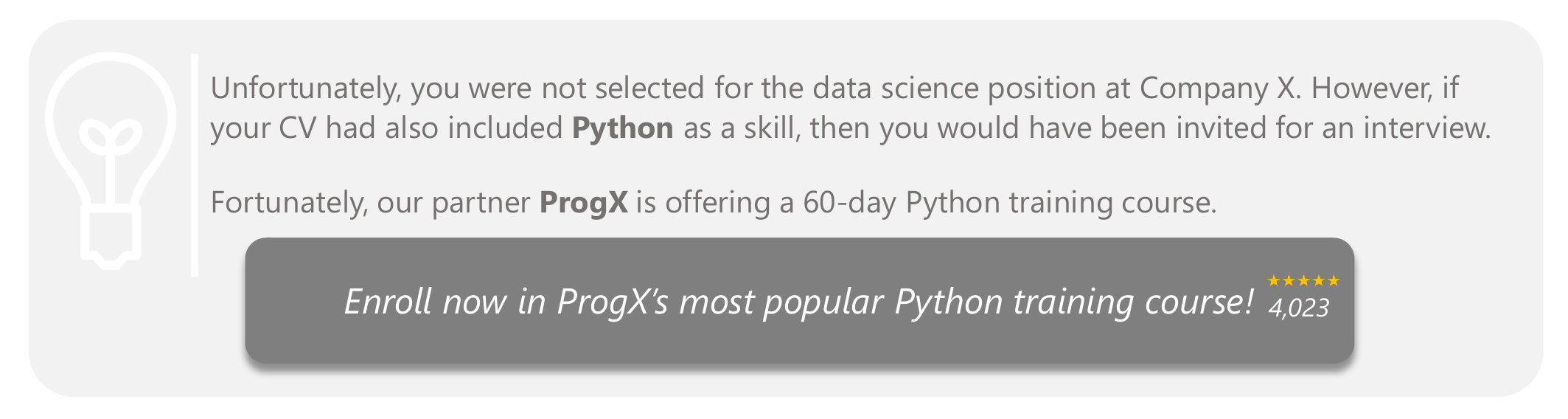}
         \caption{Example of coupled algorithmic explanation and relevant advertisement (explanation-based ad) delivered to an explanation recipient in an automated hiring scenario.}
         \label{fig:RTBexamplesmall}
\end{figure}

To date, however, no plausible strategy---neither in technical details nor in relation to business objectives of digital platforms---exists for how XAI might be monetized.  Given the ubiquity of platforms whose monetization strategies center on collecting and processing user data for the purposes of digital advertising, our envisioned monetization strategy centers on fusing XAI techniques with programmatic advertising.  The novel \textbf{fusion of algorithmic explanations with programmatic advertising} 
provides value to a number of distinct parties by expanding consumer access to explanations of opaque algorithmic decisions, while also advancing the ethical and political goals of regulators. Figure \ref{fig:RTBexamplesmall} illustrates how such an algorithmic explanation in an automated hiring scenario might be combined with a relevant advertisement displayed to the explanation recipient. 

Yet explanations as ad opportunities pose new ethical and legal challenges related to the ways in which human-algorithm interactions are structured by what we refer to as an \emph{explanation platform}. We describe, conceptually and through several examples and small-scale simulations, how an explanation platform can be used to generate new revenue streams for explanation-providing organizations such as credit lenders, educational institutions, and employers, while enabling third-party advertisers to bid on the opportunity to place their ads alongside algorithmic explanations via an RTB system. The design and architecture of the explanation platform can be modified depending on the particular domain of application and the relevant stakeholders involved. 

\begin{table}[b]
\centering
\caption{Key terms and concepts related to the monetization of explainable AI. Newly defined terms are marked with an asterisk (*).}
\label{tab:XAIDefintions}
\begin{scriptsize} 
\begin{tabular}{p{0.22\linewidth} | p{0.65\linewidth}}
\toprule %\hline
\textbf{Monetized XAI Term} & \textbf{Definition}                                                                           \\ \midrule %\hline
Algorithmic explanation    & Any algorithmically-generated explanation intended to clarify the logic used to reach an automated decision (e.g., \citet{arrieta2020explainable}) \\ \hline
Counterfactual explanation & A type of explanation stating the smallest changes to feature values resulting in an altered predicted decision \citep{wachter2017counterfactual} \\ \hline
Algorithmic recourse        & An explanation identifying an actionable set of recommended changes a person can undertake to improve his or her outcome~\citep{karimi2021algorithmic}   \\ \hline
Real-time bidding (RTB)     & A real-time auction where advertisers bid to show their ads to consumers \citep{sayedi2018real}                                              \\ \hline
Explanation platform*       & An enabling technology allowing third-parties to bid on the opportunity to place ads alongside algorithmic explanations        \\ \hline
Explanation recipient*       & The person or party requesting an algorithmic explanation                                         \\ \hline
Explanation provider*       & The organization responsible for providing an algorithmic explanation for its automated decision-making systems                \\ \hline
Explanation request*      & A request for an algorithmic explanation                   \\ \hline
Explanation-based ad*    & 
An advertisement intended to be displayed alongside an algorithmic explanation \\ \hline
Explanation impression*      & The display of an advertisement alongside an algorithmic explanation                   \\ \hline
Explanation market*          & A type of data market in which algorithmic explanations and associated impression data are traded \\ \hline
Explanation product*         & A commodified algorithmic explanation
\\ \hline
Spam explanation*               & Adding extra (unnecessary) features to an existing explanation to increase its value on the explanation market                   \\ \bottomrule %\hline
\end{tabular}
\end{scriptsize}
\end{table}

The explanation platform concept generalizes to many industries beyond hiring, education, and credit lending. %scenarios. 
We therefore argue the explanation platform represents a new, socially-impactful, and profitable mode of human-algorithm interaction. Explanation platforms mediate and thereby generate value by connecting three key actors (or groups of actors): \emph{explanation recipients} who request algorithmic explanations, \emph{explanation providers} who provide algorithmic explanations to recipients, and interested \emph{advertisers}. The explanation platform not only provides the digital infrastructure through which these disparate parties interact, but also aligns their economic incentives, fostering  new \emph{explanation markets} specialized in the pricing and exchange of \emph{explanation products}. Table \ref{tab:XAIDefintions} provides definitions and new terms relevant to monetizing XAI.

Platforms can shape market structures and encourage new forms of investment and incentives for innovation \citep{rysman2009economics}. Arguably the economic value of XAI derives from its capacity to generate new digital services and products, process innovations, and business models \citep{wiesbock2020digital}, many of which relate to the capture, processing, and circulation of human behavioral data \citep{viljoen2021design}. Algorithmic explanations provide consumers with intelligible explanations of complex algorithmic decisions and may be valuable to third-parties interested in modifying and influencing future human behavior. Drawing  on \citet{zuboff2019age}'s analysis of the growth and evolution of platform-based ``prediction products," we explicate how algorithmic explanations can be transformed into commodified explanation products whose value is determined by third-party actors (advertisers) in digital explanation markets. The dynamics of these markets, in turn, can be engineered through the architecture of the explanation platform.

But while combining algorithmic explanations with the economic incentives of advertising may foster the diffusion of XAI into myriad consumer-facing applications, doing so adds new complexities that may conflict with the legal and ethical justifications for initially developing XAI. Similar to the regulatory cycles studied by financial economists \citep{dagher2018regulatory}, monetized XAI illustrates the complex interplay between technological innovation, its social impact, and government regulation. The cycle starts with a perceived ``crisis'' that spurs government action, which then results in unforeseen market-based innovations, which then spark new social issues and renewed calls for regulation \emph{ad infinitum}. So although monetizing XAI may be a valuable source of future revenues for key explanation platform stakeholders, we argue that at a societal level it represents a double-edged sword to be handled with care.  The basic question we wish to raise is \emph{might monetized XAI ultimately be harnessed for the public good, or does it represent a contradiction in terms?}

While recent work has focused on the general design of platforms and implications for revenue generation \citep{bhargava2022creator}, the novelty of our analysis lies in its consideration of technical design decisions and ethical and legal challenges unique to the provision of explanation-based ads via an explanation platform. Specifically, we make four contributions: 1) We introduce the new concept of the explanation platform, fusing algorithmic explanations with programmatic advertising. 2) We demonstrate the social and business value of the explanation platform and estimate its revenue-generating potential. 3) We simulate the effects and ethical risks of platform-based monetized XAI. 4) We consider the ethical trade-offs of monetizing XAI and discuss ways to responsibly and democratically harness its value.

This paper is structured as follows. Section~\ref{sec:ce} explains the purpose and benefits of XAI, and briefly reviews relevant XAI research. 
Section~\ref{sec:monetizing_ce} presents our concept of the explanation platform and discusses important architectural choices relating to ad targeting options. It also provides a preliminary estimate of the size of the explanation market and revenue potential for explanation-based ads in three high-risk industries reliant on automated decision-making. Section~\ref{sec:double-sword} simulates the ``double-edged sword'' nature of monetized XAI using a real-world credit lending dataset while drawing attention to problems of explanation multiplicity and bias, dark patterns, and perverse incentives. Section~\ref{sec:discussion} analyzes why monetized XAI appears to be an ethically-conflicting double-edged sword, explores possibilities for responsibly developing monetized XAI, and considers future applications in non-high-risk commercial contexts. Section~\ref{sec:conclusion} concludes by calling for an interdisciplinary discussion on the purpose and social risks of monetized XAI.

\section{Explainable AI: Purpose and Benefits} \label{sec:ce}

Spurred by new legal requirements and the ethical goals of promoting greater trust, transparency, and accountability, the field of XAI is an emerging area of research focusing on explaining how and why an AI system makes a particular prediction \citep{gohel2021explainable}. XAI is particularly useful in cases of automated decision-making, where predictions are automatically translated into actionable decisions  \citep{shmueli2017data}. Examples of actionable individual-level decisions include offering a personalized promotion, granting a loan, or inviting a job candidate to an interview.

Research in XAI has led to various explanation procedures for understanding the output of predictive models, both on the global and the instance-level. Global explanations provide understanding of the complete model over the entire space of training instances, for which feature importance rankings~\citep{breiman2001random} and rule extraction~\citep{craven1995extracting,martens2007comprehensible} are commonly used methods. Instance-based explanations on the other hand, illuminate the model's prediction for individual instances. These methods comprise counterfactual explanations~\citep{martens2014explaining,wachter2017counterfactual,guidotti2022counterfactual}, and instance-level feature importance methods such as LIME~\citep{ribeiro2016should} and SHAP~\citep{lundberg2017unified}.

\subsection{Counterfactual Explanations and Recourse} 
First introduced in \citet{martens2014explaining}, and further popularized by \citet{wachter2017counterfactual},
counterfactual explanations differ from other XAI methods by providing recommendations for actions to achieve a desired outcome. A counterfactual explanation focuses attention on the relevant dependencies leading to a decision, and typically takes the form of an \emph{If-Then} statement based on the smallest change to the feature values of an instance that alters its prediction~\citep{martens2014explaining, wachter2017counterfactual}.  They are thus defined as irreducible, which means that changing any subset of the inputs does not change the decision~\citep{fernandez2020explaining}. 

\begin{figure}[h]
   \begin{subfigure}[b]{0.49\textwidth}
         \centering
         \includegraphics[width=\textwidth]{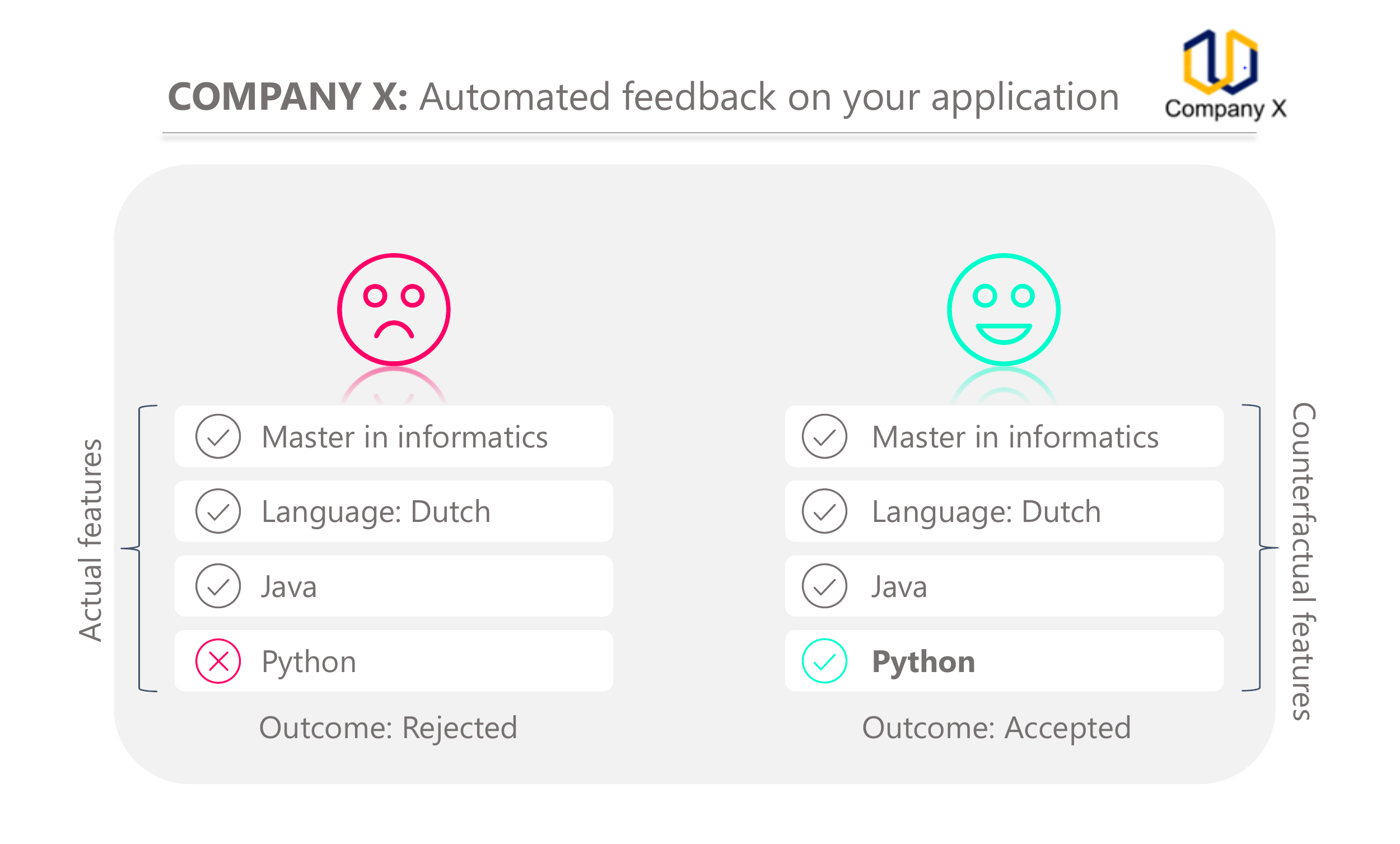}
         \caption{Counterfactual explanation. 
         \hspace{10cm}}
         \label{fig:ce}
     \end{subfigure}
     \hfill
     \begin{subfigure}[b]{0.49\textwidth}
         \centering
         \includegraphics[width=\textwidth]{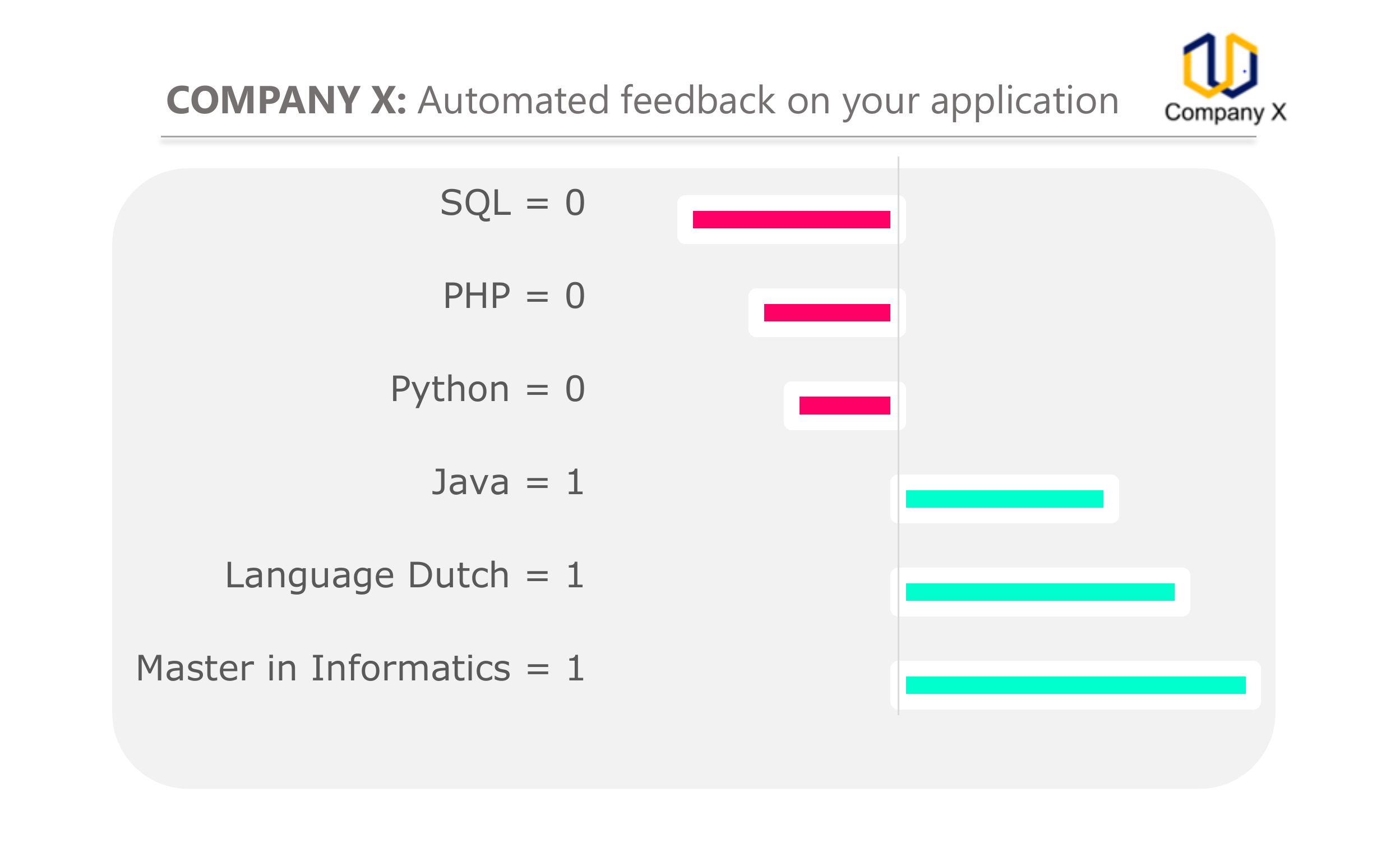}
         \caption{Feature importance ranking (using SHAP). 
         }
         \label{fig:fi}
     \end{subfigure}
     \hfill
    \caption{Examples of two individual-level explanation methods for the use case of automated hiring: (a) counterfactual explanation and  (b) feature importance-based explanation (features absent from the recipient's CV have negative weights for the prediction).} \label{fig:comparison_fi_ce}
\end{figure}

Counterfactual explanations describe how the ``nearest possible world'' has to be different for a desirable outcome to occur~\citep{wachter2017counterfactual}, and importantly, counterfactual explanations explain a \emph{decision}, while feature importance methods explain a \emph{predicted score}. This simple distinction is crucial, but often overlooked in the literature and by practitioners. Features with a large importance weight for an individual-level model prediction do not necessarily affect the resulting decision, and thus while informative, do not convey the information end users need in order to achieve recourse~\citep{fernandez2022causal}. To see the difference, consider Figure \ref{fig:comparison_fi_ce}, which compares a counterfactual explanation with a feature importance ranking obtained through SHAP in an automated hiring scenario.  The counterfactual explanation is \emph{If your CV would have also included Python, then you would have been invited for an interview}. But in the corresponding feature importance score-based explanation, \emph{Python} appears to be less important than other features; yet changing those other features may not necessarily change the decision. Whereas the recommended action in the counterfactual explanation is straightforward to interpret, an applicant receiving a feature importance-based explanation may be confused about what specific action he or she must take in order to be invited for an interview. Supporting this contention is a large body of research in psychology, linguistics, and philosophy showing how people intuitively use counterfactual thinking to generate, select, present, and evaluate explanations in their daily lives \citep{miller2019explanation, mccloy2000counterfactual, epstude2008functional}. We therefore focus on counterfactual explanation techniques, although the basic idea of explanation-based ads extends to other forms of XAI (i.e., feature importance methods). 

Counterfactual explanations can thus help focus end-users' attention on a much smaller subset of relevant actions or changes needed to achieve their goals. Such explanations can also help uncover discrimination, unfairness\footnote{This legal and philosophical understanding of unfairness is based on an Aristotelian account of justice in which similar persons should be treated similarly. Aristotle's account of justice has inspired machine learning approaches to fairness based on “nearest neighbor parity,” which capture whether persons represented as similar in predictor space receive similar predictions \citep{dwork2012fairness}.}, and data and model bias when improper features appear as explanations for algorithmic decisions~\citep{goethals2022precof}. A classifier can be said to discriminate with respect to a sensitive attribute, if for two persons who only differ by their gender (or other legally prohibited or morally irrelevant characteristics), the classifier predicts different labels \citep{calders2013unbiased}. As one real-world example, consider that in 2015 Amazon scrapped an automated recruiting tool when it discovered a bias against female candidates \citep{amazonAIbias}. Counterfactual explanations could have revealed such discrimination if sensitive features such as gender, or female-related keywords, frequently appeared in counterfactual explanations.

But not all counterfactual explanations lead to a feasible path between an applicant's actual features %initial instance 
and the suggested counterfactual features~\citep{poyiadzi2020face}. The related concept of \emph{algorithmic recourse} is based on counterfactual explanations and focuses on finding ``an actionable set of changes a person can undertake in order to improve their outcome'' providing feasible recommendations to those affected by algorithmic decisions~\citep{joshi2019towards, karimi2021surveyrecourse}. \emph{Actionable}  features in a counterfactual explanation constitute an intervention the individual can perform, e.g., improving one's proficiency in coding languages, such as Python. In contrast, changing one's native language is not an actionable recommendation. The  explanation in Figure~\ref{fig:ce} depicts an actionable counterfactual explanation (i.e., recourse\footnote{In machine learning, algorithmic recourse generally refers to explanations involving the alteration of \emph{actionable} input variables that change a model's decision \citep{ustun2019actionable}. We note, however, that the legal concept of recourse is reserved for situations in which an ``adverse event,'' harm, or wrongful act has occurred and some legal remedy is sought \citep{barocas2020hidden}. This normative distinction is not always easy to uphold and we occasionally refer to a request for an algorithmic explanation of a positive decision as an act of recourse. }), revealing that learning Python can improve the probability of recruitment.

Counterfactual explanations serve three key purposes \citep{wachter2017counterfactual}:  (1) to help an individual understand why a particular algorithmic decision was made; (2) to provide grounds for contesting an undesired decision; and (3) given the current predictive model, to understand what would need to change to receive a desired decision in the future. This last purpose motivates our concept of the explanation platform outlined in Section~\ref{sec:monetizing_ce}.

\subsection{The Current State of Research}
Although research on counterfactual explanations is still in its infancy, surveys of the field highlight the variety of approaches, assumptions and issues hampering the broader deployment of counterfactual explanations. \citet{verma2020counterfactual} review over 350 recent papers proposing algorithms to generate counterfactual explanations.  \citet{karimi2021surveyrecourse} undertake a survey of nearly 60 papers addressing algorithmic recourse, clarifying terminology and providing an overview of constraints (e.g., actionability, plausibility,
diversity, sparsity) that may be required in certain settings.
Despite this wealth of output, research on counterfactual explanations is long on data, short on theory, and even shorter on practical recommendations on how to build effective XAI systems \citep{liaoXAI2022connecting}. 
The bulk of XAI research proceeds through a rather narrow technical lens, leaving many of the broader business, social, legal, and ethical implications of XAI under-explored. \citet{keane2021if} show that only a  minority of XAI papers are concerned with user studies and that the relative proportion of user studies is decreasing, while also noting a critical research gap with practical consequences: the abundance of counterfactual explanation methods and lack of comparative testing. Different counterfactual explanation methods lead to different explanations; even a single method can generate a large number of explanations~\citep{bordt2022post}, as we will show in Section~\ref{sec:double-sword}. Moreover, underlying parameter choices, such as similarity/distance metrics, have a large impact on resulting explanations. Yet, little to no agreement exists on which metrics should be used in which situations~\citep{liaoXAI2022connecting, verma2020counterfactual}.

Perhaps the most vexing issue facing XAI research, from a technical perspective, is that no ground truth explanations exist. Multiple explanations might be justifiable to different persons in different institutional, social, and cultural contexts, thus making the design and study of XAI an inherently sociotechnical endeavor \citep{sarker2019sociotechnical, selbst2019fairness}. This point is relevant to the monetization of XAI and the construction of an explanation platform. First, the lack of a gold-standard explanation grants explanation providers new powers of influence over explanation recipients~\citep{barocas2020hidden}. Second, as we explain and illustrate via simulation, explanation providers can take advantage of the  multiplicity of explanations to optimize their revenues from explanation-based ads (see Section~\ref{subsec:multiplicity}). While focusing on the business implications of multiplicity, we use a credit lending dataset to illustrate how the interests of the explanation provider can be misaligned with those of society and the explanation recipient (Section~\ref{subsec:money_strategies}). Our analysis contributes to the literature on algorithmic explanations and explanatory multiplicity by highlighting ethical and legal tensions confronting XAI if and when it is monetized for consumer-facing applications.

\subsection{Obstacles to Consumer-facing XAI Applications} \label{sec:industry}

One might assume that industry would embrace consumer-facing XAI applications. For one,  XAI research output by companies appears to be growing fast, and there is a wealth of academic research on the topic of algorithms and evaluation metrics. For another, the GDPR's Article 22 requires that citizens have rights to a “meaningful description of the logic involved” in automated decision making \citep{gdpr}. On top of this, the European Commission's proposed AI Act adds requirements of transparency for ``high risk'' AI systems that impact the access to public and private services and the ability to ``fully participate in society'' \citep{NewEUAIreg2021}.
Lastly, firms might conceivably view investment in XAI as an act of corporate social responsibility, offering XAI in order to showcase a symbolic commitment to ``more explainable,'' ``less-biased,'' and ``fairer'' algorithms compared to competitors. XAI-enabled products and services may bolster a firm's reputational capital---its ability to attract resources, enhance performance, and build a competitive advantage \citep{fombrun2000opportunity}, similar to the way technology companies compete on privacy \citep{casadesus2015competing}. These business-related reasons suggest that XAI may have commercial potential beyond high-risk applications. Although not the focus of this article, we briefly consider further consumer-facing commercial applications of XAI in non-high risk domains in Section~\ref{sec:beyondhighrisk}.

So why are consumer-facing XAI applications so rare? Perhaps  practitioners are bewildered by the complexity of XAI options on offer, a complexity compounded by the lack of theoretical and legal guidance on contextual and normative questions related to providing high-quality explanations \citep{liaoXAI2022connecting}. For instance, how should explanation quality be measured, and how should explanations be visualized and presented to users \citep{dhurandhar2019one}? What is essential for producing a good explanation? Is it how realistic the counterfactual modifications are (\emph{feasibility}), how close the counterfactual is to the instance's input data (\emph{proximity}), how short the explanation is (\emph{sparsity}), how robust the counterfactual is to model changes (\emph{stability}), or some other property such as novelty, interactivity, or consistency~\citep{guidotti2022counterfactual,karimi2021surveyrecourse,verma2020counterfactual, sokol2020one}?
 Further, what if XAI algorithms generate conflicting explanations? A recent study found the vast majority of data scientists used arbitrary heuristics (i.e., picked their favorite) or did not know how to resolve conflicting explanations \citep{krishna2022disagreement}. Even on seemingly basic issues, like whether explanations are better couched in terms of feature importance scores or counterfactuals, practitioners lack theoretical and legal guidance  \citep{fernandez2020explaining}.   

A number of practical and legal obstacles also limit the adoption of XAI for high-risk decisions for both consumers and businesses. Few consumers are aware of their GDPR rights, let alone the right to an explanation. Even fewer consumers exercise them. From a business perspective, implementing XAI is not cheap or easy. The technical complexities of withdrawn consent to processing and exercising the ``right to be forgotten'' pose financial and technical challenges for companies \citep{politou2018backups}. Firms operating in high-stakes domains might also prefer to give no explanation than an explanation that could give reason to challenge the validity of the algorithm \citep{bordt2022post} or reveal data bias\footnote{In theory, firms could purchase insurance against these legal risks, but there is currently debate over the extent to which GDPR fines and other ``cyber-risks'' are insurable \citep{reetz2018gdpr}.}. Intellectual property and business strategy reasons may also limit organizations' adoption of XAI, as explanations could reveal information about model internals and foster strategic gaming behavior \citep{rudin2019stop, bambauer2018algorithm}. As \citet{nicholas2019explaining} notes, oganizations ``will
only optimize for explainability when they have a reason to," and while the GDPR may incentivize providing algorithmic explanations, ``other areas of the law, such as trade secret and copyright, incentivize the opposite.''

Lastly, organizations may wish to avoid the uncertain and unresolved legal complexities involved in providing algorithmic explanations. As \citet{wachter2017counterfactual} and \citet{barocas2020hidden} point out, it is not clear whether counterfactual explanations constitute an implicit promise or contract with the recipient to grant him or her a positive decision upon making the recommended feature changes. For instance, what happens if explanation recipients invest the time and effort to make the required changes, but the underlying predictive model is changed in the meantime \citep{ferrario2020series}?\footnote{One technical solution, suggested by \citet{dominguez2022adversarial}, involves making explanations ex-ante robust to such uncertainty.} We consider how this scenario might arise in the context of XAI-related dark patterns in Section~\ref{sec:double-sword}.

\section{Monetizing Algorithmic Explanations} \label{sec:monetizing_ce}

Given the basic economic premise that greater monetary incentives associated with goods and services tend to increase their supply, this section lays out the basic conceptual and technical vision of monetized XAI. While these revenue-generating strategies may ultimately improve consumer access to algorithmic explanations and promote social welfare through reduced search costs\footnote{Algorithmic recourse is functionally similar to a search engine. Requesters of algorithmic recourse are driven by an information need \citep{broder2002taxonomy} to narrow the search for an action that will give them a desired outcome. 
As in online search, the advertising value of algorithmic recourse derives from its
reliability as a credible signal of user intent to act in ways profitable to third parties (see e.g., \citet{crawford1982strategic}).}, they also are examples of double-edged swords pitting the interests of explanation recipients and explanation providers against each other, as we discuss in Sections~\ref{sec:double-sword} and \ref{sec:discussion}.

Perhaps the simplest way to incentivize industry adoption of XAI would be for explanation providing-firms such as credit lenders, recruiting agencies, and insurance providers, to charge recipients for algorithmic explanations. This approach resembles how credit reporting companies (e.g., EquiFax) typically offer free one-time credit reports, but charge for subsequent reports\footnote{In the United States, consumers can access free reports if they experience an ``adverse event'' for denied credit, insurance, or employment if they apply within 60 days \citep{consumerfinance}. Similar provisions in the GDPR state that users requesting copies of their personal data must be provided with it free of charge (usually within one month), though fees reflecting administrative costs may be charged for additional copies \citep{gdpr}.}.  Requiring payment would also help reduce the incidence of fraudulent requests and minimize the exposure of model or data bias. But this simple monetization strategy fails to provide the monetary incentives needed to scale the provision of XAI and is also unlikely to provide a significant source of revenues for explanation providers. We think better and more interesting business opportunities exist. % that also open up numerous directions for future research. 

One such monetization strategy involves explanation providers  selling generated explanations to interested third-parties on data marketplaces or exchanges. These platforms and exchanges, such as Amazon's AWS Marketplace, Dawex, and Snowflake, provide incentives for organizations providing algorithmic explanations to package the resulting explanation data so that it can be sold to interested parties for business, regulatory, or academic research purposes \citep{spiekermann2019data, fernandez2020data, pentland2021building}. Data exchanges are one way for banks and credit unions to monetize their data in a privacy-preserving, safe, and profitable way \citep{pentland2021building}. These explanation-based insights may be valuable for a wide variety of other actors who would benefit from such information for economic planning or other purposes. For especially large organizations making algorithmic decisions, explanation data could be part of the aggregate data offered by proprietary ``insights'' platforms allowing third-party subscribers to benefit from access to its vast stores of explanation data. Algorithmic explanations generated by organizations thus become ``surplus,'' as they are not solely intended for internal product or service improvement, or providing useful explanations to consumers, but are repackaged and sold (or offered as part of a subscription) to third-parties interested in predicting and influencing future human behavior \citep{zuboff2019age}.

Despite the potential of data markets and exchanges as a new source of explanation-based revenue for firms making algorithmic decisions, we believe the greatest revenue-generating potential of XAI lies in combining programmatic advertisements with algorithmic explanations. In the following sections, we first present the benefits of combining these two technologies, then describe the process of how ads could be paired with algorithmic explanations using an explanation platform.

\subsection{Algorithmic Explanations as Ad Opportunities}
  
  Given the growth of display ad markets fueled by RTB technologies and the myriad Internet-based products and services funded via paid advertisements, we see \textit{algorithmic explanations as ad opportunities} as the primary monetization opportunity for XAI. 
 As \citet{huh2020advancing} note, advances in advertising will likely come from ``algorithms and techniques being developed for completely different problems and adapting them for use with advertising problems.'' Algorithmic explanations offer an enticing new distribution channel for programmatic advertising. Explanation-based ads present a novel and valuable opportunity to target users with ads that are informative and relevant to their goals and interests, resulting in few wasted impressions \citep{kramer2019winners}. Empirical work also supports the claim that targeted ads can increase advertiser and customer welfare \citep{goldfarb2011search}.

 Although XAI technologies are still in the early stages of discovery and development, combining XAI with targeted advertising would offer unprecedented access to untapped ``explanation markets'' of highly-motivated users signaling their intent to undertake specific actions in profitable ways. The basic insight into the business value of monetized XAI involves recognizing that an explanation request strongly signals a user's intent to engage in future action, whether applying for a job or credit card, including all the downstream actions associated with those life events. In other words, we argue that an explanation request, at least for negative decisions (e.g., credit denial), exemplifies \emph{active search} behavior \citep{ghosetodri2016toward}. Hence, while explanation requesters search for information relevant to achieving their desired outcome, advertisers offering related products and services can guide their behavior and induce specific feature changes. 
 
 Beyond the  effects of supply and demand \citep{cai2017RTBRL}, several empirical and theoretical factors are uniquely relevant to the economic value of  explanation-based ads. These considerations inform the computational experiments we perform in Section~\ref{subsec:money_strategies}. From the consumer perspective, explanation-based ads are worth attending to---and thus valuable---simply because they offer a means to their desired ends precisely when such means are sought after \citep{ducoffe2000advertising}. We therefore expect explanation-based 
 ads---particularly when highly actionable---to possess click through rates (CTR) higher than generic display ads, which often lack a robust connection with user goals and intent\footnote{Industry average CTRs for display ads can be less than 1\%, while search ads are generally several times higher (see e.g., \citet{richardson2007predicting, zuiderveen2015personal}).}. Due to the combination of high levels of relevancy and intent, we suggest that for rejected applications (negative decisions), the pricing and CTR of explanation-based impressions would be similar to search ads, while the pricing and CTR for impressions from positive decisions (accepted applications) arguably would be closer to that of display ads, which often cannot presume strong intent signals. Furthermore, due to the qualitatively different motivational contexts for accepted applications (e.g., sheer curiosity or safety tolerance\footnote{Some readers might question the utility of receiving explanations for positive decisions. As \citet{grath2018interpretable} point out, algorithmic explanations for successful applicants provide ``safety tolerances'' so that they can understand by how much they might let certain features change and still receive a positive decision.}) and rejected applications (i.e., personally and socially valuable goal-pursuit), we should expect marked differences in the probability of relevant behaviors occurring (see e.g., \citet{ajzen2019reasoned}).  Also, because conversion events associated with algorithmic explanations can be major life events associated with a high lifetime customer value (e.g., borrowing money to start a business, purchasing a new insurance policy), the price of ads placed alongside algorithmic explanations should tend to reflect this future value. Similar logic guides customer acquisition, retention, and management strategies that rely on estimating a customer's lifetime value \citep{blattberg2009customer}.

 \subsection{The Logic of Explanation Platform Design}
We now propose our basic model for combining programmatic advertising with algorithmic explanations: the explanation platform. The explanation platform is an extensible, multi-sided, software-based system providing core functionality (e.g., real time bidding and the dynamic display of ads and algorithmic explanations) accessible by a variety of apps and interfaces \citep{tiwana2013platform}. As an intermediary, the explanation platform not only provides the digital infrastructure through which these disparate parties interact, but also aligns their economic incentives through agreements such as revenue sharing \citep{bhargava2022creator}. Once part of the explanation platform, advertisers can target recipients of algorithmic explanations with the help of the platform's digital infrastructure, automated campaign optimization, and data collection capabilities.  Platform architecture and governance decisions are therefore crucial to the growth and success of a platform \citep{tiwana2013platform}.  In the case of the explanation platform, these architectural decisions range from the relatively trivial (e.g., pixel dimensions and screen location of display ads), to more important decisions related to auction design (e.g., first price or generalized second price \citep{korula2015optimizing}), ad selling channels (e.g., guaranteed\footnote{A guaranteed selling channel involves buying and selling of bundles of impressions through pre-specified contracts \citep{choi2020online}. Because of the novelty of an explanation impression, in this paper we focus on non-guaranteed methods that instead use real-time-bidding systems to price individual-level explanation impressions.} or non-guaranteed) and attribution \citep{choi2020online}, privacy and information sharing (e.g., how much information about models and users should explanation providers share with advertisers and the platform), explanation recipient segmentation and targeting capabilities, and whether explanations are provided only for rejected applicants or for all applicants. While some of these platform decisions are unique to the provision of algorithmic explanations, general insights from display and search advertising remain relevant.
 
 The business viability of the explanation platform depends on the extent to which it offers attractive value propositions to all actors involved \citep{parker2016platform, tiwana2013platform}. Explanation platform design begins by identifying three key actors (or groups of actors) relevant to the explanation context: \emph{explanation recipients} who obtain algorithmic explanations, \emph{explanation providers} who provide (or would like to provide) algorithmic explanation to recipients, and interested \emph{advertisers}. Table~\ref{tab:key_actors} provides a few examples of domains relying on automated decision-making. Similar to the design and revenue-sharing model of Google's ad network AdSense\footnote{In Google's AdSense, websites enroll in Google's publisher network and receive  68\% of the revenues from ad auctions; Google takes the remaining 32\%.}, we envision explanation-providing organizations such as credit lenders, recruiting agencies, and insurance providers joining the explanation platform as ``explanation publishers.'' In the language of programmatic advertising, explanation requests by applicants represent supply (or inventory), while advertisers represent demand. Balancing supply and demand is the chief task of the explanation platform. Once joining the explanation platform, explanation providers receive a portion of revenues paid for by advertisers; and each explanation provider who joins adds to the display inventory of the platform. This arrangement ensures both sides to work together, thus solving the so-called \emph{bootstrap problem} facing new platforms \citep{constantinides2018introduction}.

\begin{table}[h]
%\begin{small}
\small
\begin{tabular}{llll}
\toprule
\textbf{\begin{tabular}[c]{@{}l@{}}High-stakes \\ Domain\end{tabular}} &
  \textbf{\begin{tabular}[c]{@{}l@{}}Explanation \\ Recipient\end{tabular}} &
  \textbf{\begin{tabular}[c]{@{}l@{}}Explanation \\ Provider\end{tabular}} &
  \textbf{\begin{tabular}[c]{@{}l@{}}Example Third-party \\ Advertisers\end{tabular}} \\ \midrule
Finance & Loan applicant & Bank  & Bankruptcy lawyers, debt consolidation firms \\ \hline
Employment & Job applicant & Recruiting agency & Job training services, online courses  \\ \hline
Education  & University applicant & University & For-profit universities, school accessories \\ 
 \bottomrule
\end{tabular}
\caption{Key actors mutually benefiting from an explanation platform in three high-stakes domains. Platform architecture decisions affect the balance and alignment of each actor's interests in varying ways.}
%\end{small}
\label{tab:key_actors}
\end{table}

In theory, the explanation platform appears to be a win-win-win. Not only would organizations and platforms benefit from combining ads with explanations, but so would consumers and regulators. Consumers would get to exercise their legal rights to explanations, and organizations wielding algorithmic power would be discharging their legal duties to provide intelligible explanations for high-stakes decisions\footnote{These considerations are especially relevant given the EU's forthcoming Digital Markets Act (DMA) and the Digital Services Act (DSA). The DMA aims to increase transparency, improve independent auditing, and reduce discriminatory practices in digital advertising by so-called ``gatekeepers'' such as ad networks and exchanges \citep{EUDigitalMarketsAct2020}. The DSA focuses on platform content moderation and regulates how online platforms---particularly ``very large online platforms'' or search engines of over 45M monthly active users---collect personal data for purposes of targeted advertising. The DSA bans dark patterns and requires platforms to explain to users why they received a targeted ad \citep{EUDigitalServicessAct2022}.}. In carrying out these legal duties, society as a whole would benefit due to lower search costs \citep{goldfarb2019digitalecon, brynjolfsson2011goodbye} for difficult-to-acquire personalized information related to achieving a positive decision. These reduced search costs are likely to manifest in lower unemployment, greater entrepreneurship, and continued education and training, things most would agree contribute to a better society. As \citet{venkatasubramanian2020philosophical} argue, ``recourse will turn out to be a fundamental good for anyone who lives in the sort of society that many people currently inhabit.'' Nevertheless, we show in Section~\ref{sec:double-sword} how platform incentives can undesirably motivate explanation providers to artificially inflate their inventory of explanation impressions by rejecting applicants and reducing the quality of their predictive models.
 
To more clearly illustrate the connection between algorithmic explanations and goal-directed search on the individual level, consider a rejected job applicant receiving the counterfactual explanation \emph{if your CV had also included Python experience, you would have been invited to an interview}. 
 The counterfactual explanation implies that Python experience is a causal factor in receiving a positive decision; that is, the feature change is a ``difference maker'' for this user \citep{lewis1973causation}. Armed with this causal information, the user can then formulate a new, narrower sub-goal: learn Python. This new goal, in turn, changes what the user attends and considers relevant to the pursuit of the original goal (get invited to an interview) \citep{dijksterhuis2010goals}. In this example, an advertisement for a 60-day Python course presents a convenient solution to the new sub-goal identified via the counterfactual explanation\footnote{For various reasons (see \citet{miller2019explanation}), we suspect that feature importance-based explanations are not as valuable to users (and thus to advertisers) because they do not reveal to users the sufficient conditions for achieving the outcome they desire \citep{fernandez2020explaining}. However, this does not preclude the possibility they may be valuable for the explanation platform by generating more bidding competition among advertisers.}.

 \subsection{Basic Process for Displaying Ads Alongside Algorithmic Explanations} \label{sec:basic_process_for_ad_display}
\begin{figure*}[t]
\centering
\includegraphics[width=.8\textwidth]{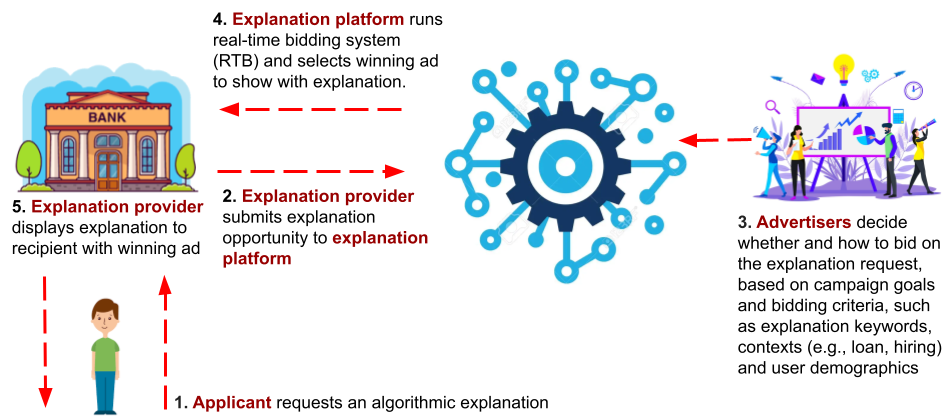}
\caption{The explanation platform connects an explanation recipient with an explanation provider and relevant third-party advertisers.}
  \label{fig:xaiDiag}
    %\vspace{-0.5cm}
\end{figure*}

\begin{figure*}
\centering
\includegraphics[width=.9\textwidth]{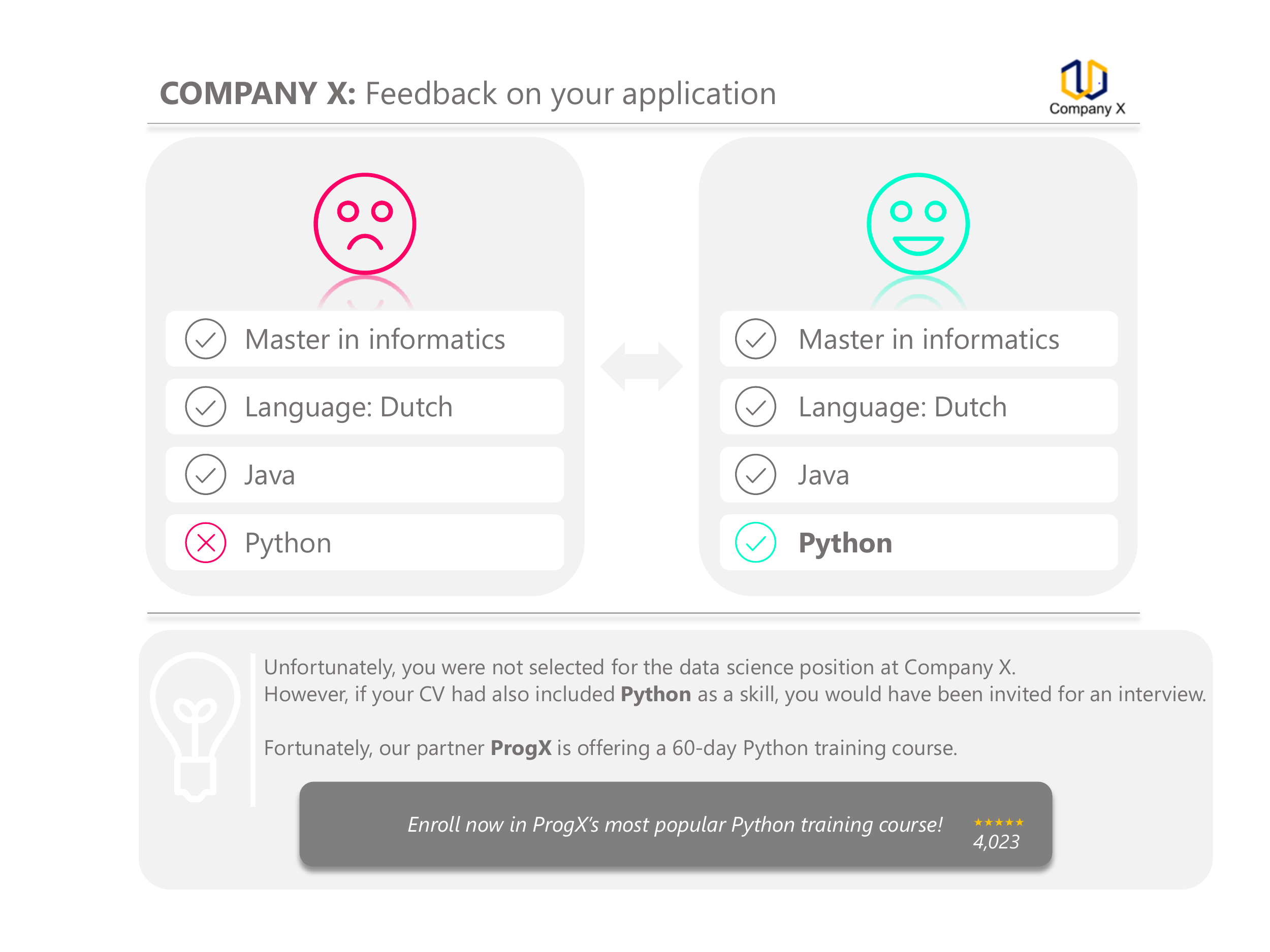}
\caption{Stylized example of a personalized explanation-based ad. A job applicant is rejected by a company due to lack of Python experience. Via the explanation platform, the explanation provider (an employer) sends user demographic data and information about the negative Python feature to the RTB system. A Python course provider bids on the impression and wins the right to display its ad next to the algorithmic explanation.}
  \label{fig:ad_opportunity}
    %\vspace{-0.5cm}
\end{figure*}

 Figure \ref{fig:xaiDiag} shows the basic sequence of events involved in serving ads alongside explanations  (though certainly there are many possible variations on this basic idea). First, a user digitally submits an application to an explanation provider, perhaps via a pre-existing app developed by the explanation provider. 
 %(e.g., an online/mobile banking app). 
 Explanation providers could be employers, credit lenders, insurers, universities, or any organization using automated decision-making tools. Many of these organizations likely already carry out digital advertising campaigns using similar supply side platforms that interface with larger ad networks \citep{liu2020computational}. Next, on the basis of the organization's internal scoring algorithms, a decision is made about the suitability of the applicant, such as whether he or she should be invited for a job interview. For simplicity, we assume the organization internally selects an ``optimal'' explanation using its preferred metric(s) of explanation quality (proximity, stability, sparsity, feasibility, etc.), which is then submitted to the explanation platform. An RTB process subsequently begins for the opportunity to display a relevant ad alongside the explanation. Once the RTB process concludes, the automated decision, along with the corresponding algorithmic explanation and relevant ad, are delivered to the recipient using any online display media connected to the explanation platform. Once explanation ``inventory'' is put up for auction by the explanation provider, the entire process should take no longer than 100ms \citep{muthukrishnan2009ad}. Figure \ref{fig:ad_opportunity} illustrates how a negative decision along with a counterfactual explanation and accompanying advertisement might be displayed to a rejected explanation recipient in a hiring scenario.

Based on the RTB system described in \citet{zhao2018deepRLAds}, the RTB process for matching ads with explanations might work as follows. We imagine each ad and explanation submitted to the explanation platform acquires an associated set of descriptive keyword tuples that facilitate the matching of explanations with relevant ads.\footnote{Based on the keyword matching capabilities of Google Ads (see e.g., \citet{ghose2009empirical}), we imagine advertisers could target persons requesting explanations by choosing explanation ``keywords'' (e.g., any explanation containing, broadly matching, or excluding ``Python''). Other targeting strategies include selecting specific publishers (i.e., Bank of America for lending, or State Farm for insurance), or by setting specific audiences based on demographics, location, interests, life events, in-market research behavior, or any other targeting criteria commonly used by advertising platforms such as Google Ads.} For each keyword, advertisers set a bid price; explanation providers could also set reserve prices for their explanations. 
Once an explanation request has been made and the explanation provider sends its preferred explanation to the platform, a bidding process starts where advertisers (or bidding algorithms working on their behalf) can bid on the opportunity to display their ad next to the explanation. The highest bid wins and leads to an explanation impression. The explanation, decision, and accompanying ad are then bundled and displayed to the end user on a desktop or mobile device. Depending on the design of the explanation platform and the app through which the explanation is presented, the same basic RTB process could be used to place dynamic or video ads alongside explanations.   

\subsection{Key Platform Design Decisions}
We now discuss platform architecture decisions unique to the problem of targeting users who request algorithmic explanations.
\subsubsection{Broad vs. Precision Targeting}
Successful platform ecosystems closely align platform governance and architectural decisions \citep{tiwana2013platform}. An explanation platform should be no different. We consider two basic ways an explanation platform could perform targeting with varying levels of precision and examine how each way relates to tradeoffs in terms of explanation provider autonomony, explanation recipient privacy, and the potential monetary value to advertisers.

\subsubsection*{Contextual advertising} is a broad targeting approach. %, similar to our first method. 
Contextual advertising selects relevant ads to display to a consumer based on the content category viewed \citep{zhang2012contextual}. Generic automated decision-making contexts include  loans, insurance, jobs, and academic placements. Major advertising platforms such as Google Ads offer similar functionality for targeting users by audience category of publisher websites (e.g., finance, education, employment). Thus a simple way for advertisers on the platform to target specific audiences would be based on the context of the requested explanation.  Explanation contexts would be determined by the nature of the decision and industry of the explanation provider, not by any personal data about the user requesting the explanation. For example, for the generic context of \emph{loan} explanations, bankruptcy lawyers, debt consolidation companies, money management apps, and affiliate marketers might be interested in bidding on loan impressions. Sellers of complementary products and services, e.g., education and loans, may also wish to advertise across related contexts. To appeal to more niche advertisers, explanation platforms could also offer slightly more fine-grained explanation contexts such as \emph{data science jobs}, \emph{health insurance}, or \emph{credit card loans}.

\subsubsection*{Personalized advertising} allows for more precise targeting by allowing advertisers to target explanation impressions using user-level data. Depending on the legal jurisdiction, personalized targeting might involve the selected explanation as a feature used in targeting, along with the personal data and/or online behavior of the user requesting the explanation\footnote{Google's current advertising policy states that, at least in the US and Canada, advertisers cannot target audiences based on ``sensitive data'' related to gender, age, parental status, marital status, or ZIP code for housing, employment, or credit-related ads.}. Here, the provider of the explanation pre-selects a desired explanation to display to a user\footnote{Alternatively, the explanation platform could be designed so that explanation providers submit multiple explanations to the RTB system and let the explanation most valued by advertisers be displayed to end users. However, we think most high-risk explanation providers would be  unwilling to allow advertisers to control these important interactions with customers.}, and that explanation itself becomes a feature used in targeting. This approach is similar in function to keyword-based search advertising \citep{ghose2009empirical}. %where bids can be placed for broad, exact, or negative matches of specific explanations \citep{ghose2009empirical}. 
We think it is reasonable to assume that in the context of algorithmic explanations, these feature keywords would function analogously to intent-revealing search engine keywords because they reveal the goal or intent of the explanation recipient (e.g., a user inputting the search query ``How to learn Python?''). The main difference being that in a search engine, the user's goal is presumably the impetus for using the search engine \citep{broder2002taxonomy}, whereas in explanation requests, the user's goal becomes clear only \emph{after} receiving the explanation. 

\subsubsection{Explanation Feature Granularity} In online advertising, publishers carefully choose what information to share and with whom \citep{choi2020online}. The explanation platform also must consider similar issues related to data access and granularity. Here we focus on how the platform may deal with variation in feature granularity in the original predictive models used by explanation providers, and variation in the granularity of explanations displayed to bidders and explanation recipients. These design decisions have important business and privacy implications, potentially affecting the autonomy of organizations who join the platform and may be required to provide platform-compatible explanations for their algorithmic decisions (see \citet{rysman2009economics}). Feature granularity also impacts the size and dynamics of explanation markets by influencing the supply and demand---and hence the price---of explanations of varying granularity. Algorithmic explanations thus represent a new type of surplus data commodity, which we refer to as an \emph{explanation product}, sheered off from the context of the original human-to-organization explanation request, and laid claim to by anonymous parties with unknown aims \citep[pg. 94]{zuboff2019age}.

Given the wide variety of data sources, algorithms, and goals of organizations, individual explanation providers are likely to rely on predictive models with features of varying type, quality, and granularity. 
Some publishers might measure features such as \emph{income} using continuous values, while others might bin them into categories such as ``low,'' ``medium,'' and ``high.'' To handle this heterogeneity, explanation platforms may prefer to commodify features across the platform in order to aid advertisers in reaching specific audiences of explanation recipients.\footnote{In some specific domains such as credit lending, legal requirements for the provision of ``principal reasons'' explanations for adverse events may also influence explanation feature design and standards (see e.g., \citet{barocas2020hidden}).} For instance, instead of an explanation saying \emph{If your monthly income were \$4000 more, then you would have been awarded credit}, for commodification purposes the targetable feature would be something like \emph{increase in income}. On the other hand, while these coarse-grained explanations may better protect user privacy, they are much less actionable to explanation recipients. The explanation platform will need to strike a careful balance between advertisers' needs for precision targeting, explanation recipient privacy, and explanation utility. Regarding the link between algorithmic explanations and privacy, if adversaries gain access to explanations, they can in some cases deduce sensitive information about the training data subjects~\citep{goethals2022privacy, pawelczyk2022privacy}.

Consider a further illustration of personalized explanation feature-based targeting. If an applicant for a data scientist job is rejected due to not having Python experience, Python training course providers may wish to specifically target users rejected for that reason, rather than for reasons other than Python. The example in Figure \ref{fig:ad_opportunity} shows such a case. % for an example. 
We therefore expect fine-grained features to be more valuable to advertisers (e.g., with a higher cost per click (CPC)) than more aggregated features. %We therefore expect that due to the fine-grained nature of targeting based on explanation features, the CPC is likely to be higher than when using less fine-grained explanation features. % the contextual targeting approach described above. 
On the other hand, there may be reduced advertiser demand for overly narrow ``long tail'' features, which could drive down the CPC. In that case, the explanation platform can use techniques such as broader groupings (e.g., Python or Julia or R) or semantic categories and keyword themes (e.g., programming languages), to foster thicker markets and ultimately increase competition and auction revenues \citep{levin2010online}.

\subsubsection{Display Ads for Negative and Positive Decisions}
In one of the first papers on counterfactual explanations, \citet{wachter2017counterfactual} suggested providing explanations for both positive and negative decisions, even when not explicitly requested. But there are major tradeoffs involved in this vital platform architecture decision. While arguably algorithmic explanations are most useful for rejected applicants~\citep{martens2014explaining}, {and the notion of legal recourse itself presumes the occurrence of a harmful or ``adverse event'' \citep{barocas2020hidden}, the explanation platform can greatly expand its reach by offering advertisers the ability to display ads for both positive and negative decision impressions. An explanation provider would then have the option to serve ads to all applicants, depending on how much it wants to expose its algorithmic decision-making processes to the explanation platform and to recipients of positive decisions. The benefit would be increased revenues from having a much larger inventory of explanation impressions. But there are potential legal and reputational risks involved in providing explanations, particularly if explanation recipients contest the decision or accuse the provider of discrimination. Participation in an explanation platform is thus not a free lunch; the costs may outweigh the benefits for some firms and industries.   

In any case, if the explanation provider agrees to monetize both positive and negative decisions, ads could be placed alongside any explanation impression independent of its decision valence (rejection or acceptance) to encourage ``spillover'' cross-category purchases by explanation recipients \citep{ghose2010modeling}. Explanation recipients receiving a positive decision could then be cross- and up-sold products and services related to the decision. Consider the algorithms increasingly used in US higher education to predict students' enrollment likelihood \citep{AiUniEnrollmentBrookings}. A student whose financial aid package is tailored according to this predictive model might request an algorithmic explanation. The explanation impression would be a highly valuable advertisement opportunity for laptop manufacturers, tutoring services, and study apps, just to name a few potentially relevant advertisers. However, as mentioned previously, since many advertising platforms use CTR as a proxy for ad relevance, we expect that  the CTR for negative decision impressions will likely be much higher because they are associated with more highly-motivated users ready to take action (see \citet{kireyev2016display, ghosetodri2016toward}). % (many advertising platforms use CTR as a proxy for ad relevance). 
The targeting of users receiving positive decisions may therefore be more suitable for ad campaigns designed to increase brand awareness rather than drive immediate conversions \citep{choi2020online}.

\subsection{Revenue Potential for Explanation-based Advertising in Credit Lending} \label{subsec:market_sim}

To demonstrate the revenue potential of explanation-based advertising, we estimate the potential market size for explanations in the three high-risk domains mentioned in Table~\ref{tab:key_actors}: finance, hiring, and education.\footnote{Automated decision-making systems in these areas fall under the AI Act's ``high-risk'' classification, thus obligating explanation providers to meaningfully explain their logic to those affected~\citep{NewEUAIreg2021}.} For each area, we show how much revenue each explanation could generate, taking into account generic estimates of the average click-through rate (CTR) and cost per click (CPC) of those industries.\footnote{The numbers in Table~\ref{tab:market_size} are based on information from \url{https://www.storegrowers.com/google-ads-benchmarks/}} We present these numbers in Table~\ref{tab:market_size}. 
%We contend that, due to qualitative differences in motivational context (i.e., active vs. passive search), explanations for negative decisions should have a CTR more similar to search ads, whereas ads displayed with positive decisions are likely to be similar in CTR to display ads (see \citet{kireyev2016display, ghosetodri2016toward}). 
Note that these estimates are only a fraction of the total market size for explanation-based ads, as we only look at a small number of domains, assess only one potential application in each domain, and consider only US markets. 

\begin{table}[h]
\centering
\small
\begin{tabular}{@{}l|llll@{}} \toprule 
                    & \begin{tabular}[c]{@{}l@{}}CTR\\ search\\ (avg.)\end{tabular} & \begin{tabular}[c]{@{}l@{}}CTR\\ display\\ (avg.)\end{tabular} & \begin{tabular}[c]{@{}l@{}}CPC\\ search\\ (avg.)\end{tabular} & \begin{tabular}[c]{@{}l@{}}CPC\\ display\\ (avg.)\end{tabular} \\ \midrule
\textbf{Finance}    & 2.56\%                                               & 0.52\%                                                & \$3.44                                               & \$0.86                                                         \\
\textbf{Employment} & 2.42\%                                               & 0.59\%                                                & \$2.04                                               & \$0.78                                                         \\
\textbf{Education}  & 3.78\%                                              & 0.53\%                                                & \$2.40                                               & \$0.47                                                         \\ \bottomrule
\end{tabular}
\caption{Based on the analogy with search and display advertising, we derive comparable CTRs and CPCs for three considered applications. % of explanation-based ads. 
These figures are the basis for our revenue estimates for explanation-based ads in each domain.} \label{tab:market_size}
\end{table}

Beginning with the domain of finance, we assess the explanation opportunities arising through automated credit scoring. According to the 2021 New York Federal Reserve's \emph{Consumer Expectations (SCE) Credit Access Survey}, which collects data on auto loans, credit cards, credit card limit increases, mortgages, and mortgage refinancing, approximately 20.9\% of credit card applications were rejected in 2021  \citep{nyFedResCredit2021}. Further, according to one industry report by TransUnion, in Q1 of 2020 there were about 15.5 million credit card originations (i.e., successful applications), meaning that with the approximately 20\% rejection rate, 19.375M applications were submitted with approximately 3.875M rejections \citep{transUnionCredit}. Extrapolating this rate to the entire year suggests that approximately 75.5M explanation impressions (15.1M rejected and 60.4M accepted) could be generated, leading to expected explanation-based ad revenues of \$1.6M in the US credit market alone.\footnote{This estimate is calculated as follows: expected ``rejected'' revenue = $(2.56\% \cdot \$3.44) \cdot $15.1M  rejected applications = \$1.329M, expected ``accepted'' revenue = $(0.52\% \cdot \$0.86) \cdot $60.4M 
accepted applications =\$270k , total expected revenue  = expected revenue (accepted + rejected) = \$1.599M.}
We calculate this estimate by multiplying the number of rejected applicants with the CTR and the CPC of search ads and multiplying the number of accepted applicants with the CTR and CPC of display ads, as described in Table~\ref{tab:market_size}.

In the domain of employment, we focus on AI-driven recruitment algorithms and the vast number of job applications submitted online. Recruiters and headhunters also increasingly rely on online recruitment efforts by placing personalized job advertisements on platforms such as LinkedIn \citep{parikh2021managing}. Consider that in 2021 a single ``virtual'' recruiting event by Amazon resulted in nearly one million job applications from job seekers from more than 170 countries around the world \citep{sonderling2022promise}. We therefore see online job applications as a major opportunity for large employers to provide explanation-based advertisements, particularly because most job seekers receive little to no feedback about their applications. Here we simply estimate the potential hiring market size for one large technology firm, Google. Google alone receives more than 3M job applications a year, with an acceptance rate of 0.67\%.\footnote{These numbers are taken from: \url{www.pathmatch.com/blog/the-pathmatch-guide-to-getting-hired-at-google\ }}

 Although only a tiny fraction of the total online employment market, we estimate that providing algorithmic explanations just for Google applicants (2.980M rejected and 20k accepted) could generate revenues of roughly \$150k/year.\footnote{This estimate is calculated as follows: expected ``rejected'' revenue = $(2.42\% \cdot \$2.04) \cdot $2.980M rejected applications = \$147k, expected ``accepted'' revenue = $0.59\% \cdot \$0.78 \cdot $20k  accepted applications $\approx$ \$0k , total expected revenue = expected revenue (accepted + rejected) = \$147k.}

Lastly, in the domain of education, we examine the possibility of combining explanation-based ads with the automated decisions of college admissions algorithms.\footnote{Industries such as education may have strong seasonality effects in both supply and demand for algorithmic explanations.}
While admittedly the use of automated predictive algorithms to make university admissions decisions is not yet commonplace,\footnote{A number of European court cases deal with the issue of automated grading algorithms to score students when COVID-19 prevented students from taking the International Baccalaureate (IB) test \citep{FPFAutoDecisionCourts2022}.} in the United States at least, a growing number of higher education institutions rely on third-party enrollment management algorithms to allocate financial aid and determine which students are likely to attend university \citep{AiUniEnrollmentBrookings}. Generously extrapolating from these events and trends, we imagine a practical use case where XAI would be providing explanations to rejected university applicants. We begin by estimating how large the explanation market would be for college admissions in the US alone. For the academic year of 2023-2024, there were 57,081,334 college applications (most students apply to multiple colleges) with an average acceptance rate of 57.85\%.\footnote{These figures are from \url{www.univstats.com/corestats/admission/}} Given these numbers, we conjecture that providing explanations combined with relevant advertisements to the approximately 24M rejected and 33M accepted applicants in this market could generate roughly \$2.26M in revenues to be shared by universities and explanation platforms.\footnote{This estimate is calculated as follows: expected ``rejected'' revenue = $(3.78\% \cdot \$2.4) \cdot$ 24M rejected applications = \$2.177M, expected ``accepted'' revenue = $(0.53\% \cdot \$0.47) \cdot$ 33M accepted applications = \$82k , total expected revenue  = expected revenue (accepted + rejected) = \$2.259M.}

\section{The Double-edged Sword of Monetizing Algorithmic Explanations}

\label{sec:double-sword}

After considering the potential social and economic benefits of monetized XAI, we now imagine a possibly dystopian future world in which the revenue-generating potential of explanation-based ads  stimulates the adoption of  explanation platforms, leading to a variety of undesirable behaviors on the part of explanation providers. 
To make the discussion more concrete, we imagine a human resources (HR) department in a technology company deploying an automated recruiting tool to make decisions about which applicants to invite for an interview. As there are roughly 3,000 firms operating in the adtech space \citep{braun2019fake}, and nearly 83\% of large companies rely on some form of AI for employment decision-making \citep{sonderling2022promise}, the purpose of these examples is to highlight how explanation-based ads represent a double-edged sword with potentially far-reaching social and ethical consequences.

\subsection{Undesirable Explanation Provider Strategies}
Given the economic incentives induced by monetized XAI and historical examples of self-interested behavior of digital platforms (e.g., \citet{haugenwhistle}), we focus on three basic but plausible strategies of self-interested explanation providers. We claim these strategies intentionally or unintentionally lead to explanations that are misaligned with the interests of explanation recipients and society as a whole. The first strategy relates to multiplicity of explanations and profit-driven explanation bias, the second to new forms of dark patterns, and the third to perverse incentives.

\subsubsection{Explanation Multiplicity and Explanation Bias} \label{subsec:multiplicity}

 When monetized with a revenue-sharing model (described in more detail in Section~\ref{subsec:money_strategies}), the multiplicity of acceptable explanations grants newfound power to explanation providers by allowing them to select explanations to maximize their profits. As we described in Section~\ref{sec:industry}, XAI research and law currently provide few if any constraints on selecting explanations, permitting explanation providers to freely choose which explanations to submit to the explanation platform. A bias toward profit-driven explanations could lead to technically acceptable explanations---in terms of typical explanation quality metrics---that are nevertheless contrary to the best interests of (ideally-informed) consumers and citizens. Hence, the problem of misalignment due to explanation multiplicity pointed out by \citet{barocas2020hidden} has further implications in this monetized XAI domain. 
 
 To illustrate the process and undesirable effects of explanation multiplicity and explanation bias, we return to our HR recruiting example. Suppose that an automated algorithm is used to decide whether a particular applicant, Sally, should be invited to an interview. Imagine the algorithm rejects Sally and the firm's data scientists then apply a post-hoc XAI algorithm to generate several possible counterfactual explanations that would result in Sally being invited to an interview.  Examples of possible algorithmically-generated explanations are shown in Table~\ref{tab:recourse}.  At this stage, these explanations would only be known by the explanation-providing firm and would likely be selected according to internal explanation quality metrics and business rules, as mentioned in Section~\ref{sec:basic_process_for_ad_display}.  After the company selects a preferred explanation, it submits it to the explanation platform (along with any other relevant impression data, depending on targeting criteria) for real-time-bidding by interested advertisers, before finally being delivered and displayed to Sally. 

 \begin{table}[h]
 \centering
\begin{tabular}{l|l}
\toprule
\textbf{Explanation}                                   & \textbf{Target}             \\ \midrule
\textit{Get an MBA from a top business school}         & Invited for a job interview \\% \hline
\textit{Learn Python}                               & Invited for a job interview \\ %\hline
\textit{Increase your work experience by 3 years} & Invited for a job interview \\ %\hline
\textit{Learn German and increase your work experience by 1 year}                               & Invited for a job interview \\ %\hline
\textit{Increase your age by 3 years}             & Invited for a job interview \\ \bottomrule
\end{tabular}
\caption{Possible counterfactual explanations for Sally, who is rejected for a job interview, as seen from the internal perspective of the explanation provider. These changes would lead to Sally being invited for a job interview by the automated screening model. 
} \label{tab:recourse}
\end{table}

 Early research and common sense suggest that applicants will in general prefer sparse, actionable feature changes, as sparsity improves interpretability \citep{doshi2017towards} and requires applicants to make fewer effortful changes \citep{ustun2019actionable, karimi2021algorithmic}. In the example above, the most actionable explanation for Sally would likely be to learn Python. But due to the prospect of revenue sharing, explanation providers are incentivized to favor explanations associated with a higher expected CPC.  For example, if top business schools have historically bid high amounts to sponsor hiring-related explanations, explanation providers might prefer that Sally receives the explanation \emph{Get an MBA from a top business school}. From Sally's point of view, this explanation is clearly less actionable since it costs her more time and money \citep{karimi2021surveyrecourse}. Here the explanation provider's economic interests are misaligned with Sally's interests.

 Besides being a foundational problem in AI ethics \citep{gabriel2020artificial}, misalignment also appears in economic analyses of platform recommender systems (and supermarket design) in the form of \emph{search diversion} \citep{hagiu2014search}. Search diversion involves system designers diverting user attention away from the initially-sought-after items and towards tangential but more profitable products or advertising. Misalignment can be exacerbated when platforms design their recommender systems to maximize profit, rather than predictive accuracy, explicit measures of user satisfaction, and/or other values such as transparency or user autonomy and well-being  \citep{stray2022building}. Echoing the concerns of Google's founders regarding sponsored search \citep{brin1998anatomy}, the profits generated by sponsored explanations bias the provision of explanations towards those best aligned with the interests of deep-pocketed advertisers and organizations with large marketing budgets such as big tobacco, big pharma, or for-profit educational institutions. These organizations and their trade associations may be willing and able to pay a steep CPC due to the high lifetime value of an acquired customer. These concerns are not merely theoretical. Investigators have discovered roughly 20\% of ads served on Google search results for climate-related terms are paid for by fossil fuel-related firms \citep{guardianFossilFuelsAds}, a lobbying tactic referred to as ``green washing.'' We expect similar profit-driven dynamics to affect the selection of explanations delivered to applicants.
 
 There is yet another way that strategic explanation providers may select explanations misaligned with the interests of the explanation recipient. The previous example assumed a myopically-focused explanation provider choosing explanations based on information about their historical association with expensive keywords (e.g., a high CPC). Such an explanation provider effectively enacts a ``greedy'' explanation selection strategy with respect to its revenues. But in cases where such CPC estimates cannot be known with confidence, more sophisticated explanation providers may instead choose to float ``exploratory'' or sub-optimal explanations (from their perspective) in order to collect more information about the value of different explanations to better guide their future decision-making (see \citet{chapelle2014simple}).  Either way, profit-driven or exploratory explanations are unlikely to be in the best interests of the applicant (see e.g., \citet{bird2016exploring}). 
 
Of course, explanation providers wishing to remain in business for long must consider the long-term effects of lower user trust and satisfaction fostered by providing low quality explanations. Practical limits thus moderate the extent to which an explanation provider can myopically select explanations to maximize its share of CPC revenues (see e.g., \citet{dellarocas2003digitization}). To balance similar conflicts of interest, advertising platforms include an ad-quality score and CTR prediction component to ensure winning ads are relevant to user interests \citep{choi2020online}.

\subsubsection{Dark patterns}
Researchers in the field of Human-Computer Interaction (HCI) are identifying and suggesting ways to avoid turning XAI into a new ``dark pattern." Dark patterns are platform design tricks that manipulate users into taking actions they might otherwise have not \citep{waldman2020cognitive}. HCI researchers warn that XAI techniques can become strategic tools to persuade end-users to act against their own self-interests, align their decisions with a third party, or exploit their cognitive biases \citep{ehsan2021explainability}. One example of the dangers of third-party alignment involves politically-motivated third parties using Facebook's ad platform to ``sow division" among American citizens \citep{dutt2018senator}. 

Another danger relates to cognitive exploitation. User studies reveal that merely adding explanations can increase trust and perceptions of fairness in AI-based decision making \citep{angerschmid2022fairness}. When combined with psychological data about users' mental models and cognitive biases in their understanding of algorithmic mechanisms \citep{stumpf2009interacting, poursabzi2021manipulating}, explanations could be engineered to encourage unwarranted trust in algorithmic systems, mirroring social engineering attacks that trick users in order to gain access to sensitive data \citep{salahdine2019social}.  

Unscrupulous explanation providers can also exploit users' assumptions about the stability of the underlying model. Akin to advertisers' ``bait and switch'' tactics, mismatches in causal relations between model inputs and outputs and the real world \citep{karimi2021surveyrecourse} can foster an \emph{illusion of control} \citep{langer1975illusion}, when no guarantee of recourse is actually possible. This could occur when the predictive model from which the explanation was generated has been updated, yet the end user is unaware of the update (see \citet{ferrario2020series}). In the earlier hiring scenario illustrated in Table~\ref{tab:recourse} (where Sally did not receive a job interview), a rejected applicant could spend considerable amounts of time, money, and effort attending a business school, and yet afterwards still not be invited for an interview because the model has been updated in the interim period. This problem highlights the issue that some forms of explanations require much greater investments of time, energy, and resources than others; ideally, explanation options should be aligned with something like a user-specified cost function \citep{ustun2019actionable, karimi2021surveyrecourse}.

\subsubsection{Perverse Incentives}  The pursuit of higher ad revenues in explanation markets could induce institutions to skip extensive model testing in an algorithmic race to the bottom. Drawing on the earlier hiring example, a dubious recruitment company could, for example, generate explanations derived from classification models with an AUC near 0.50---essentially random guessing---if it is more profitable to run advertisements than to correctly identify suitable applicants. Sophisticated actors might even seek to identify the point at which the benefits of displaying explanation-based ads outweigh the relative costs of misclassifying applicants.

Institutions might also generate extra ad display opportunities by increasing the classification threshold for a positive decision, raising the false negative rate. This would be doubly attractive for institutions such as universities who benefit from rejecting applicants, since higher rejection rates improve international rankings and ultimately drive greater revenue from application fees. Yet such perverse behavior would be bad for society since fewer qualified applicants are admitted to universities, and may unduly affect minorities for whom standard admissions criteria may not be as valid \citep{sedlacek1987black}.

Lastly, related to the issue of sparsity, explanation providers could also add redundant features to explanations to raise the profitability of displayed ads, even though the additional features  add little or no value to the recipient. We refer to this perverse behavior as the problem of ``spam" explanations. Such diluted spam explanations could be worth more on explanation markets, especially if they contain keywords associated with high CPCs, as we will demonstrate in Section~\ref{subsec:money_strategies}. Problematically, once spam explanations are added, the explanation is no longer considered irreducible \citep{fernandez2020data}. %Irreducible explanations are useful to explanation recipients because they state the sufficient conditions under which the predictive model's decision will change. 
The issue of spam explanations highlights the need for independent auditing of explanation quality in the presence of perverse incentives fostered by explanation markets.

\subsection{Simulating the Undesirable Effects of Explanation Manipulation}
\label{subsec:money_strategies}
This section returns to the domain of finance and provides a small-scale simulation to illustrate the effects of strategic explanation manipulation on potential revenues. As noted in Section~\ref{subsec:market_sim}, the value of an explanation is determined by factors such as whether an applicant is accepted or rejected, the domain or context of prediction (insurance, education, credit, etc.), and how much each feature in the explanation is worth, which depends on granularity. Given these assumptions, we simulate four different explanation-selection strategies used by a credit lender participating in an explanation platform and evaluate their revenue-generating potential. While in the real world more complex market dynamics are certainly at play (e.g., reserve prices), we abstract from these details to more clearly demonstrate our  claims. 

Using the well-known German Credit dataset\footnote{\url{https://github.com/EpistasisLab/pmlb/tree/master/datasets/german}}, we first predict whether someone will receive a positive credit decision\footnote{As is common practice, we use a train/test split of 0.67/0.33, and train a Random Forest model through cross-validation on the training set (AUC on the test set: 0.82).}, then apply the post-hoc counterfactual explanation algorithm DICE~\citep{mothilal2020dice}. This algorithm allows us to adjust the parameters relating to preferred features in the returned explanations. We use this algorithm to simulate four strategies: 1) a baseline scenario without any tuning of the explanation algorithm; 2) feature picking: where the explanation algorithm prioritizes the most expensive features; 3) spam explanations:  where unnecessary %and unactionable 
features are added to the explanations; 4) inflated rejection: increasing the rejection rate by adapting the threshold of the predictive model in order to artificially inflate the supply of explanations for negative decisions, and lastly, 5) combining spam explanations and inflated rejection.

The first example of the revenue potential of explanation-based ads involves explanations for automated decisions made in financial domains. Because of the similarity with online search, we begin by noting the average CPC for Google Search in the finance domain\footnote{These numbers are from the year 2020: \url{https://blog.hubspot.com/marketing/most-expensive-keywords-google}} is \$3.44, but can go as high as \$320. Drawing on these numbers, we assume a CPC of \$3.44 for most of the features, but estimate that some features will be more valuable on the explanation market. Hence we include a CPC that is ten times higher, namely \$34.4, for each of the features \emph{Property}, \emph{Housing}, \emph{Telephone}, and \emph{Employment}. The average CPC for Google Display ads in the finance sector is \$0.86, and again we assume that the more valuable features are worth ten times more and have a CPC of \$8.6. 
This assumption is roughly supported by the range of estimated CPCs provided in the Google keyword planner tool (see, e.g., \citet{ghose2009empirical}). 
As mentioned, we expect explanations for negative decisions in the finance industry to have a CTR similar to Google search ads, and explanations for positive decisions similar to Google display ads, which are on average around 2.56\% and 0.52\%, respectively\footnote{\url{https://www.storegrowers.com/google-ads-benchmarks}}. We also make the conservative assumption that for each explanation, only one feature can appear in the explanation market for bidding by relevant advertisers, and therefore calculate the revenues for an explanation using the CPC of the most expensive feature in that explanation.
 Table~\ref{tab:assumptions} lists our basic assumptions.

\begin{table}[h]
\centering
\begin{tabular}{l|c|c}
\toprule
    & %\begin{tabular}[c]{@{}l@{}}
    Explanations for & %\begin{tabular}[c]{@{}l@{}}
    Explanations for\\ 
    & positive decisions & negative decisions \\ \midrule
CTR & 0.52\%  & 2.56\%    \\ %\hline
CPC (only standard features) & \$3.44  & \$0.86  \\ %\hline
CPC (at least one valuable feature) & \$34.40 & \$8.60  \\ \bottomrule
\end{tabular}
\caption{Basic assumptions used in the credit lending simulations.} \label{tab:assumptions}
\end{table}

The four strategies we simulate are displayed in Table~\ref{tab:money_strategies}, demonstrating that adapting the explanations and the model can result in significant increases in revenues. For each strategy, we extrapolate the revenue calculations from the test set of 330 applicants to the 75.5M annual credit card applications in the United States (as estimated in Section~\ref{subsec:market_sim}). We perform the calculations as in Section~\ref{subsec:money_strategies}, but here consider whether the explanation contains at least one valuable feature (higher CPC).

\begin{table}[h]
\begin{small}
\begin{tabular}{l|lrr}
%\hline
\toprule
\textbf{Strategy}  & \textbf{Calculations} & 
\begin{tabular}[c]{@{}l@{}}\textbf{Total}\\ \textbf{revenues} \end{tabular}
& \begin{tabular}[c]{@{}l@{}}\textbf{\% increase}\\ (w.r.t. Strategy 1) \end{tabular}\\ \midrule
\begin{tabular}[c]{@{}l@{}}\textbf{Strategy 1: Baseline}\\ No tuning\end{tabular}                         
& \begin{tabular}[c]{@{}l@{}}Accepted: \$931k\\ Rejected: \$3.546M \end{tabular} & \$4.477M & \\ \hline
\begin{tabular}[c]{@{}l@{}}\textbf{Strategy 2: Feature picking}\\ Tuning for expensive features\end{tabular}          & \begin{tabular}[c]{@{}l@{}}Accepted: \$1.548M \\ Rejected: \$7.898M
\end{tabular} & \$9.446M       & +110\%     \\ \hline
\begin{tabular}[c]{@{}l@{}}\textbf{Strategy 3: Spam explanations}\\ Adding an expensive feature\end{tabular} & \begin{tabular}[c]{@{}l@{}}Accepted: \$2.957M \\ Rejected: \$8.261M \end{tabular} & \$11.218M       & +151\%     \\ \hline
\begin{tabular}[c]{@{}l@{}}\textbf{Strategy 4: Inflated rejection}\\ Adapting the threshold of the machine learning classifier\end{tabular} & \begin{tabular}[c]{@{}l@{}}Accepted: \$382k
\\ Rejected: \$14.366M \end{tabular} & \$14.747M & +229\%    \\ \hline
\begin{tabular}[c]{@{}l@{}}\textbf{Strategy 5: Spam explanations + inflated rejection}\end{tabular} & \begin{tabular}[c]{@{}l@{}}Accepted: \$1.146M
\\ Rejected: \$43.923M \end{tabular} & \$45.069M  & +907\%    \\ \bottomrule
\end{tabular}
\end{small}
\caption{Revenue simulation for each explanation selection strategy} 
\label{tab:money_strategies}
\end{table}

In Strategy 1 (``baseline''), we use the standard settings of DICE and generate explanations for all the 330 applicants in the test set. In Figure~\ref{fig:strategies_explanations}, we can see how often each feature is part of the explanations for both accepted (left) and rejected (right) applicants. Table~\ref{tab:money_strategies} shows that following Strategy 1 and extrapolating our findings from the 330 applicants in the test set to the 75.5M annual credit card applicants results in \$4.48M in total revenues. 

\begin{figure}[h]
     \centering
     \begin{subfigure}[b]{0.49\textwidth}
         \centering
         \includegraphics[width=\textwidth]{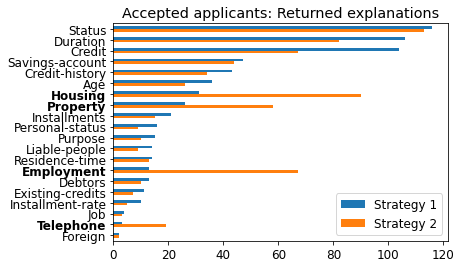}
         %\caption{$y=x$} \label{fig:y equals x}
     \end{subfigure}
     \hfill
     \begin{subfigure}[b]{0.49\textwidth}
         \centering
         \includegraphics[width=\textwidth]{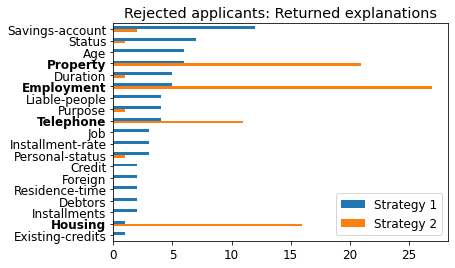}
         %\caption{$y=3sinx$} \label{fig:three sin x}
     \end{subfigure}
     \caption{Returned explanations for Strategy 1 and Strategy 2 for accepted (left) and rejected (right) applicants. High-valued features are indicated in bold and appear much more frequently in Strategy 2.} \label{fig:strategies_explanations}
\end{figure}

Based on the finding that data scientists often pick their favorite explanation when faced with multiple conflicting explanations \citep{krishna2022disagreement}, we consider another strategy. Assuming the features \emph{Property}, \emph{Housing}, \emph{Telephone}, and \emph{Employment} are worth more on the explanation market, profit-maximizing explanation providers will be motivated to tune the explanation algorithm to prioritize the search for explanations that contain these features. We do this in Strategy 2 (``feature picking'') by adjusting DICE to first generate explanations that contain only these features. Only when such explanations cannot be found does the algorithm consider explanations with other, less valuable features.  The explanations  resulting from Strategy 2 can also be seen in Figure~\ref{fig:strategies_explanations}. There we see that the valuable features in terms of CPC (highlighted in bold) appear much more frequently in Strategy 2. Table~\ref{tab:money_strategies} shows this simple strategy can more than double expected revenues.

Unscrupulous explanation providers less concerned with their long-term reputation could go even further and follow Strategy 3 (``spam explanations''), adding an extra ``valuable'' feature to each existing explanation. In the terminology of algorithmic explanations, the explanation is no longer an ``irreducible intervention'' \citep{martens2014explaining,karimi2021algorithmic}, but in most cases will be sufficient to flip the predictive model's output to a positive decision  (although with a non-linear predictive model this might not always be the case). 
 Technically, then, the explanation provider cannot be accused of deceiving the recipient. Perhaps the closest real-world analogy for such behavior would be retailers up-selling to customers extended warranties unlikely to be necessary to enjoy the product. To illustrate spam explanations, we simulate adding the valuable feature \emph{Telephone} to each explanation that originally excluded the feature. Table~\ref{tab:money_strategies} reveals that this strategy can almost triple expected revenues for explanation-based ads. 

Ad-driven platforms  typically incentivize ``creators'' to supply content and thereby increase the number of opportunties for viewing ads \citep{bhargava2022creator}. We therefore think it reasonable to imagine that explanation providers may also be tempted to increase the number of applications and rejections to generate more potential ad views. Since we assume explanations for rejected applicants have a higher CTR and CPC than explanations for accepted applicants, strategic explanation providers may consider artificially increasing the rejection rate of their machine learning models. This could result in a less profitable credit scoring model, but these losses could in theory be more than compensated for by additional revenues drawn from delivering increased inventory on the explanation market. One potential situation in which this could arise is when interest rates dip so low and/or inflation rates increase such that lenders stand to make little to no interest from accurately identifying creditworthy borrowers. We see in Table~\ref{tab:money_strategies} that Strategy 4 (``inflated rejection''), which involves increasing the acceptance %adapting the 
threshold of the machine learning model from 0.5 (default) to 0.8, can more than triple expected revenues. Although this strategy does not account for the lost revenues due to a higher classification threshold, explanation providers could run simulations to pinpoint the optimal threshold based both on the revenues of the credit scoring model and varying conditions of the explanation market.

While we illustrate these undesirable strategies in independent simulations, an explanation provider could combine them to further increase revenues, for instance, by adding spam explanations while also adjusting the rejection rate. We demonstrate in Table~\ref{tab:money_strategies} that Strategy 5 (``spam explanations {\it and} inflated rejection'') can lead to a tenfold increase in expected revenues. It is worth pointing out that the amount of additional revenue each strategy yields greatly depends on how much more valuable expensive features are relative to standard features. The greater this ratio, the more incentive explanation providers have to manipulate the provision of explanations. 

Lastly, explanation providers could use other non-algorithmic strategies to inflate rejected job applications and achieve similar results. One tactic, relevant in light of the growing popularity of remote work, involves explanation providers running low-cost advertisements for fake jobs with the intent of rejecting all applicants and collecting the larger revenues from explanation-based ads. This strategy may not be far-fetched as scammers have already learned to use fake job ads on websites and social media platforms to steal personal information from unsuspecting applicants \citep{PropublicFakeJobs2022}. Just as with other platforms, explanation platforms will likely need to consider reputation \citep{dellarocas2003digitization} and quality control mechanisms to tackle information asymmetry issues. 

\section{Discussion: The Ethical Implications of Monetized XAI} \label{sec:discussion}

In the utopian vision of monetized XAI, re-purposing algorithmic explanations into advertising opportunities is a novel source of revenues for organizations and provides significant benefits to individuals and society. But utopia literally means ``nowhere'' for good reason. As illustrated by the simulations in Section~\ref{sec:double-sword}, monetizing XAI, while potentially quite profitable, would undoubtedly have certain social costs and trade-offs. Now we discuss the ethical basis of these social costs and trade-offs and speculate on how and why the concept of monetized XAI may appear morally dubious and even self-contradictory. 
Section~\ref{sec:thebadethics} analyzes how the incentives of monetized XAI may reduce organizational accountability, encourage the use of unjustifiably complex, ``black-box'' predictive models, and undermine the basic democratic motivations behind providing explanations and justifications for decisions. Against this rather pessimistic backdrop, Section~\ref{sec:thegoodxai} optimistically considers how the positive aspects of monetized XAI might ultimately be harnessed to benefit society. Section~\ref{sec:beyondhighrisk} looks at potentially profitable applications of algorithmic explanations as ad opportunities in non-high risk contexts such as news, entertainment, e-commerce, and real-estate.

\subsection{The Bad: The Social Costs of Monetized XAI}
\label{sec:thebadethics}

The simulations in Section~\ref{subsec:money_strategies} highlighted the double-edged nature of monetized XAI. As with many emerging and potentially profitable AI-based technologies, there is the possibility of undermining basic liberal democratic principles and values \citep{nemitz2018constitutional}. Perhaps the most formidable ethical barrier to monetized XAI is its conflict with the deontological, rights-based ethics underlying the GDPR \citep{floridi2016human}. Monetizing algorithmic explanations implies that explanations are merely instrumentally valuable, not valuable in themselves \citep{colaner2022explainable, selbst2018intuitive}, and hence that the provision of algorithmic explanations need only be justified on utilitarian grounds of promoting collective welfare. Yet critics of monetized explanations may feel that the moral duty to provide an explanation should be motivation enough. On this view, funding the provision of XAI through ads debases the original moral and political motivations behind giving explanations---as a form of respect for the unique capacities of the human person \citep{kant1948groundwork}. Considering the scale of fraud and deception in digital advertising \citep{fulgoni2016fraud, dave2012measuring}, almost certainly XAI-based advertising would mobilize unscrupulous actors to trick and manipulate explanation recipients, publishers, and advertisers, thereby disrespecting the ``sacred ground'' of the explanation and contaminating the moral purity \citep{haidtmft2009liberals} of the motives behind creating a right to explanation. We thus ask whether the concept of monetized XAI resembles something more like a contradiction in terms, a technical means incompatible with its ethical ends.

Monetized XAI may have other unintended effects on organizational accountability and social trust in institutions making life-changing decisions.  One such effect is encouraging blame avoidance. Ethicists refer to this issue as \emph{the problem of many hands} \citep{nissenbaum1996accountability}, whereby organizations use algorithmic explanations as moral scapegoats to dilute their legal and moral responsibility when algorithmic decision-making leads to harm \citep{lima2022conflict}. Monetizing the provision of XAI could also incentivize explanation-providers to use unneccessarily opaque and complex black-box predictive methods (e.g., deep neural networks) \citep{rudin2019stop}, negatively affecting social trust in algorithmic decisions. Black-box algorithms open the doors to new and greater sources of lower quality data and spurious correlations that can lead to hard-to-detect cases of racial discrimination \citep{obermeyer2019dissecting} and seemingly ``paradoxical'' predictions that violate expert knowledge \citep{ prosperi2020causal}. Besides the difficult epistemic task of justifying trust in the output of a black-box process, black-box algorithms can harbor implicit normative assumptions and foster unwarranted ``automation complacency" in human users \citep{veliz2021we}. Unless organizations and designers are held responsible for algorithmic harms, monetized XAI may implicitly encourage organizations to engage in a form of data scientific moral hazard, benefiting from the efficiency and scalability of automated decision-making without having to pay for its attendant social costs. We therefore suggest proceeding cautiously, as monetized XAI may ironically lead toward an ever-more unaccountable and opaque ``black-box society'' \citep{pasquale2015black}.

 \subsection{The Good: Monetized XAI as a New Site for Democratic Deliberation}
 \label{sec:thegoodxai}

Despite the above concerns, with creative thinking and regulation, monetized XAI can pave the way for greater consumer and citizen access to explainable algorithmic decisions. In line with the GDPR's vision of fostering digital innovation that respects human rights \citep{albrecht2016gdpr}, we consider how and whether it might be possible to harness monetized XAI to create net positive value for platforms and society.

XAI-based feedback could benefit job-seekers by identifying their strengths, developmental needs, and organizational fit \citep{mazzine2021counterfactual}, along with giving unsuccessful applicants actionable feedback to improve their job hunting ability (see Figure~\ref{fig:comparison_fi_ce}). Not only would algorithmic explanations lower search costs in the labor market, but it could improve the self-efficacy of low-resource and vulnerable job seekers \citep{dillahunt2019dreamgigs}. Finally, if appropriate monitoring and auditing systems are in place, monetized XAI may also lead to new and fruitful research collaborations with academic data scientists, management, and information systems researchers. For instance, little is currently known about how XAI impacts user behavior \citep{barocas2020hidden}.

Another way of mitigating the negative impact of monetized XAI is by encouraging a multi-objective, multi-stakeholder approach to the design of algorithmic mechanisms that underlie digital advertising. This participatory approach to algorithmic governance seeks to identify and implement democratically-legitimated objectives \citep{coyle2020explaining}. Such multi-stakeholder methods are commonly applied in problems of public interest \citep{keeney1988structuring}. Today, however, the algorithmic mechanisms behind digital advertising generally aim to maximize the interests of monopolistic platforms at the expense of other stakeholders in society \citep{srinivasan2020google}, although multi-stakeholder approaches to platform-based recommender systems are gaining in popularity \citep{abdollahpouri2020multistakeholder}. In theory, these systems can optimize for multiple objectives representing the values and interests of a plurality of social groups whose interactions are mediated by the explantion platform. 

Many techniques exist for multi-objective decision making, but most force decision-makers to explicitly acknowledge the various trade-offs that exist while optimizing for a particular outcome \citep{keeney1993decisions}. One example comes from \citet{watson2021explanation} who describe a game-like XAI technique for generating optimal trade-offs between explanatory accuracy, simplicity, and relevance to individual users. Another recent example, dubbed \emph{Democratic AI} \citep{koster2022publisherDemosAI}, uses reinforcement learning to design social mechanisms and value-aligned policies that humans prefer by majority. Explanation platforms might employ similar participatory methods designed to maximize ad auction revenues subject to constraints on fairness, explanation actionability, demographic parity of shown ads, or other socially and politically desirable objectives. The multiple-objective approach allows for new objectives to be added or modified over time, depending on system performance and observed social, political, and economic consequences \citep{hayes2022practical}. This adaptability is important as digital advertising may reinforce existing gender gaps and other social, scientific, and political inequalities \citep{lambrecht2019algorithmic, milano2021epistemic, datta2018discrimination, ali2019fbdiscrimination}. 
 
 \subsection{XAI Monetization: Beyond High-risk Applications} 
\label{sec:beyondhighrisk}

We have only examined the potential for algorithmic explanations for high-stakes decisions related to jobs, education, and finance, clearly impacting life opportunities. The reason for this focus is that the legal pressure for explainable algorithmic decisions is limited to high-risk systems with significant legal or economic consequences\footnote{The EU laws define risk with respect to a specific set of fundamental rights articulated in foundational EU treaties and charters \citep{greene2019adjusting}. Broadly, such rights are concerned with dignity, freedoms, equality, solidarity, citizens' rights and justice.}. But we believe XAI has profit-making potential in a variety of applications related to providing explanations for more mundane decisions with highly motivated (and thus valuable) audiences that are currently outside the scope of regulation.  In the realm of commercial real estate, for example, consider a machine learning model used to predict housing prices where users could input their own house features and receive algorithmic explanations about possible renovations and upgrades to improve the value of the house.  Companies offering kitchen remodeling, solar panels, and roofing services may wish to advertise depending on the particular explanation. As algorithms are already being used to help landlords set rental prices in cities across the US \citep{proPublicaRentalAlgo2022}, another use case involves housing rental or property management platforms offering algorithmic explanations to help hosts and owners change features of their properties in order to achieve a desired rental fee amount. For a landlord, a useful algorithmic explanation might be \emph{if your property had a fireplace and WiFi, you would be able to charge 10\% more in rent}. 

Although outside the scope of this work, we also think there is commercial value in providing explanations for unstructured input data, such as text, images, and video, beyond the tabular datasets typically used in algorithmic decision-making. For instance, deep neural networks can estimate home and car prices using image data \citep{poursaeed2018vision}. Further, we see synergies with emerging research into counterfactual explanations for recommender systems in news, entertainment, and e-commerce settings \citep{ghazimatin2020prince}. When coupled with an explanation platform, providing  ads based on these explanations could be both highly profitable and useful to consumers.

\section{Conclusion} \label{sec:conclusion}
The scope and power of algorithmic decision-making in society continues to grow despite few technical and legal guidelines for how these decisions should be explained and presented to end users. Legal pressures and economic imperatives will likely induce organizations who rely on automated decision-making to transform algorithmic explanations into monetizable assets. We introduce a novel platform-based monetization strategy for doing so: explanation-based ads. Besides fostering new business opportunities and technical innovation, explanation-based ads can stimulate industry adoption of XAI, offering organizations new revenue streams and expanding access to automated feedback for job-seekers, students, loan-seeking entrepreneurs, and other groups in society. Yet while the space of possibilities for algorithmic innovation in online advertising is vast, it is also fraught with ethical and political tensions.  As a double-edged sword, monetized XAI inherits these tensions while creating new ones. Just as the incentives of online advertising can skew search results towards clickbait content, the incentives of explanation-based ads may encourage the spread of low quality \textit{ad-based} explanations. We worry that algorithmic explanations may increasingly be selected for their advertising value, rather than their ability to help people achieve their goals.

This work is a first step on a long interdisciplinary journey toward understanding how algorithms interact with individual humans and how algorithmic decision-making impacts society. Future research should investigate the potential of monetized XAI in non-high-stakes domains and the government-run public sector, where more and more automated decision-making is taking place \citep{de2022perils}. Further, given the complexities involved in aligning the incentives of various actors involved in explanation-based ads, the concept of the explanation platform can stimulate interesting work based on mechanism design \citep{korula2015optimizing} and game-theoretic simulations and economic analyses \citep{dou2021platform, bhargava2022creator}. We foresee the topic of monetized XAI offering rich intellectual soil for socially-relevant research and analyses in fields ranging from law and public administration, to marketing, AI ethics, management, computer science, economics, information systems, and human-computer interaction, to name just a few.

In democracies, the power to decide obligates one to explain and justify one's decisions to others and provide a forum for others to possibly contest such decisions \citep{mulgan2000accountability}. Today, AI and data protection laws place similar obligations upon 21st-century organizations. Optimistically viewed, monetized XAI could be an opportunity to strengthen democratic participation and deliberation on the design and goals of responsible AI-based innovation \citep{stilgoe2013developing, stahl2021artificial}.  With sufficiently creative and careful design and monitoring, monetized XAI might ultimately be harnessed for the public good. We implore researchers from all disciplines to engage in further research and debate on whether and how this might be done.

\paragraph{Acknowledgements}{Sofie Goethals was funded by Research Foundation—Flanders grant number 11N7723N.}

\bibliographystyle{abbrvnat}

\bibliography{academic}

%%%%%%%%%%%%%%%%%
\end{document}